\documentclass[pre,twocolumn]{revtex4-1}

\usepackage{graphicx}
\usepackage{dcolumn}
\usepackage{bm}

\usepackage[utf8]{inputenc}
\usepackage[T1]{fontenc}
\usepackage{mathptmx}

\usepackage{amsmath}
\usepackage{mathtools}
\usepackage{amssymb}

\usepackage{pst-node}
\usepackage{verbatim}   
\usepackage{color}      
\usepackage{subfigure}  
\usepackage{hyperref}   

\usepackage{braket}

\newcommand{\ba}{\begin{eqnarray}}
\newcommand{\ea}{\end{eqnarray}}
\newcommand{\be}{\begin{equation}}
\newcommand{\ee}{\end{equation}}

\pdfoutput=1

\begin{document}

\title{
Search by Return: Stochastic Resetting in Fluctuating Harmonic Potentials
}

\author{Derek Frydel}
\affiliation{Department of Chemistry, Universidad Técnica Federico Santa María, Campus San Joaquin, Santiago, Chile}

\date{\today}

\begin{abstract}
\small{ 

We study a class of stochastic resetting (SR) processes in which a diffusing particle alternates 
between free motion and confinement by an externally controlled potential.  
When the particle is recaptured, it undergoes a return trajectory that drives it toward a designated
reset point.  
In standard SR, such returns are treated as instantaneous, but in realistic setups they have finite 
duration and introduce imprecision in the starting points of subsequent search attempts.
We analyze a fluctuating harmonic potential in which return trajectories are forcibly terminated 
the moment the particle reaches the origin, ensuring that all outward (diffusive) trajectories 
begin from the same point.  
This is implemented through instantaneous positional information: a feedback signal that shortens 
the return phase without incurring additional mechanical energetic cost.  
We examine several search protocols built on this controlled return mechanism and determine their 
mean first-passage times (MFPTs).  
Of particular interest is a protocol in which outward diffusion is eliminated entirely and the 
return motion itself becomes the search mechanism.  
This “search by return’’ perspective reverses the conventional logic of SR and yields a closed-form 
MFPT that highlights the efficiency of using return dynamics as the primary search strategy.

} 
\end{abstract}

\pacs{
}

\maketitle

\noindent\textbf{
Since its modern formulation, stochastic resetting has drawn growing interest across disciplines—
as both an efficient search strategy and a nonequilibrium process. In its canonical form, it describes 
a random walk that is interrupted and a walker instantly returned to a fixed starting point. Experimentally, 
this behavior can be approximated using a fluctuating trapping potential that intermittently releases 
and recaptures a particle. However, such implementations inevitably introduce return trajectories 
and imprecision in the resetting position. In this work, we examine how to control 
these return trajectories to ensure that the initial position of a subsequent outgoing trajetory is the same. 
We then explore a range of 
search protocols that incorporate return trajectories into the search process.
}

\section{Introduction}

Stochastic resetting (SR) describes a random walk that is intermittently interrupted and 
returned to its initial position.  
Although the idea of restarting stochastic dynamics has appeared in several
contexts—including search algorithms in computer science~\cite{IPL-Luby-1993,PRL-Montanari-2002} 
and intermittency models in chemical physics~\cite{PRL-Benichou-2005,PNAS-Lomholt-2008}—
the modern statistical-physics formulation is due to Evans and Majumdar~\cite{PRL-Evans-2011}, 
who introduced diffusion with Poissonian resets and uncovered its nonequilibrium stationary
state and optimal-resetting properties.
Since then, SR has been extended to diverse stochastic processes, confinement geometries,
and search strategies~\cite{JPA-Evans-2013,PRE-Whitehouse-2013,JPA-Evans-2014,PRL-Gupta-2014,
JPA-Durang-2014,PRL-Kusmierz-2014,PRE-Pal-2015,PRE-Majumdar-2015,JPA-Christou-2015,JPA-Pal-2016,
PRL-Reuveni-2016,JSTAT-Falcao-2017,PRL-PalReuveni-2017,JPA-Pal-2022}, with comprehensive
overviews in~\cite{JPA-Evans-2020,FP-Gupta-2022,JPA-Pal-2022,Pal2024}.
Experimental realizations using holographic optical tweezers~\cite{JPC-TalFriedman-2020,PRL-Vatash-2025}
have further motivated the study of resetting in realistic potentials, especially where finite return
times and positional uncertainty become unavoidable.

A natural physical realization of SR is a fluctuating trap that intermittently releases and recaptures 
a particle~\cite{JPC-TalFriedman-2020,PRR-Besga-2020,PRL-Vatash-2025}.  
This model has attracted interest both as a variant of SR and as a nonequilibrium setup in its own
right~\cite{PRR-Besga-2020,JOPA-Gupta-2020,JSTAT-Gupta-2021,JOP-Santra-2021,JOPA-Alston-2022,
PRE-Schehr-2024,PRE-Frydel-2024,JSTAT-Mukherjee-2025}.  
However, if interpreted strictly as SR, fluctuating traps introduce a structural complication:
\emph{return trajectories}.  
These returns extend the duration of a reset cycle and introduce imprecision in the starting point
of subsequent outgoing trajectories.

The present work investigates a fluctuating trap model in which return trajectories are terminated 
at the exact moment the particle first reaches the origin, $x=0$.  
This enforces a sharp initial condition for outgoing trajectories and converts the return dynamics
from a liability into a useful ingredient of the search process.  
We also explore several search protocols that explicitly incorporate return trajectories, thereby
revealing how their structure can be exploited rather than suppressed.

The idea of enforcing return termination at the origin was first introduced in~\cite{JOPA-Gupta-2020}
to bring fluctuating-trap models closer to standard SR.  
A detailed analysis for a linear potential followed in~\cite{JSTAT-Gupta-2021}.  
Return trajectories as active components of the search protocol were subsequently explored in~\cite{PRE-Arnab-2024},
with further development for a linear trap in~\cite{PRE-Arnab-2025}.

Here we focus on a fluctuating \emph{harmonic} trap.  
Harmonic potentials are the most widely used and experimentally accessible means of implementing
controlled SR: optical tweezers, magnetic trapping, and feedback-controlled colloidal setups naturally
realize quadratic confinement with tunable stiffness $K(t)$~\cite{Pesce2020,Kumar2018,Ghosh2019,
Prohm2013,Wurger2010,Dobnikar2008}.  
Because the resulting dynamics reduces to the Ornstein--Uhlenbeck process, relaxation and first-passage
properties remain analytically tractable and directly relevant to experiments.

Return trajectories also motivate a conceptual shift.  
In standard SR, the search occurs during outward diffusion from the origin, while returns are treated as
instantaneous.  
Once return trajectories are made explicit, this logic can be inverted: {the return motion itself can 
serve as the search mechanism}.  
This “search by return’’ viewpoint is not a special case but a conceptual dual of conventional SR, highlighting
that a reset cycle contains two dynamical stages—outgoing and return—either of which may carry the search
depending on how the system is controlled.

This paper is organized as follows.  
Sec.~\ref{sec:sec1} reviews the fluctuating-trap model and the theoretical tools used throughout.
Sec.~\ref{sec:sec5} applies them to four search protocols, each exploiting return trajectories differently.
Final remarks appear in Sec.~\ref{sec:sec6}.

\section{The model}
\label{sec:sec1}

This work considers a Brownian particle in a harmonic potential that is intermittently switched on and off.  
Fluctuations of the potential prevent a particle from reaching equilibrium, giving rise to a nonequilibrium setup.  
We refer to the system with the potential turned off as the \emph{off} state, and to the system with the potential 
turned on as the \emph{on} state.

Trajectories in the "off" state are called \emph{outgoing trajectories}, while those in the "on" state are referred to 
as \emph{return trajectories}.  The overdamped Langevin dynamics governing the motion in each state is given by
\be
\dot{x}(t)=
\begin{cases}
\sqrt{2D}\,\eta(t), & \text{(outgoing)},\\[6pt]
-\mu K\,x(t) + \sqrt{2D}\,\eta(t), & \text{(return)},
\end{cases}
\ee
where \(\mu\) is the mobility, \(K\) the trap stiffness, \(D\) the diffusion constant, and \(\eta(t)\) is a zero-mean,  
delta-correlated Gaussian white noise with \(\langle \eta(t)\eta(t') \rangle = \delta(t - t')\).

The duration of each outgoing trajectory is drawn from an exponential (memoryless) distribution,
\begin{equation}
r(t)  \equiv r_{\text{off}}(t) = \frac{1}{\tau}\, e^{-t/\tau},
\label{eq:r-exp}
\end{equation}
which corresponds to a constant-rate process. As a result, the dynamics is Markovian, despite being irreversible.

Once the potential is switched on, the particle begins to drift toward the trap center. The potential remains on until 
the particle reaches the origin, \(x = 0\), for the first time. This return mechanism defines 
the end of the "on" state, and there is no predefined distribution for the return duration \(r_{\text{on}}(t)\). 
Instead, \(r_{\text{on}}(t)\) emerges from the first-passage time distribution of a Brownian particle in a harmonic trap.

This setup ensures that all outgoing trajectories restart from the position at the origin. Physically, such a 
mechanism could be realized by an ``off-switch'' located at the origin, which deactivates the potential upon the 
particle’s arrival, or by a feedback control system that instantaneously resets the dynamics upon observing the 
particle at \(x = 0\).

\subsection{steady-state distributions}
\label{sec:sec2}

The Fokker–Planck formulation of the current model has been analyzed in~\cite{JOPA-Gupta-2020}.  
To avoid redundancy, we adopt here an alternative presentation aligned with the goals of this paper 
and obtain only the quantities that will be directly used in formulating the mean first passate time (MFPT).  

At any time, an ensemble of realizations can be viewed as composed of two subensembles:
those evolving while the trap is on and those evolving while the trap is off.  
Accordingly, $n_{\mathrm{on}}(x,t)$ denotes the probability density of finding the particle 
at position $x$ at time $t$ {given that the trap is on},  
and $n_{\mathrm{off}}(x,t)$ denotes the corresponding density {given that the trap is off}.  
These do not represent different particles but the two dynamical “modes” of the same particle.
We next obtain explicit formulas for those distributions under steady-state conditions.  

Since outgoing trajectories are associated with free Brownian motion starting from the origin, 
the time-dependent probability density they give rise to is the standard Gaussian kernel:
\(
p(x,t) = \frac{1}{\sqrt{4\pi D t}}\, e^{-x^2/(4Dt)}.
\)
Because outgoing trajectories do not evolve indefinitely but are interrupted (or renewed), 
the distirbution we need to consider is \(S_r(t)\, p(x,t)\), where 
\(
S_r(t) = \int_t^\infty dt'\, r(t')
\)
is the survival function.  
The steady-state distribution due to outgoing particles is then obtained by averaging \(S_r(t)\, p(x,t)\) over time, 
\begin{equation}
n_{\text{off}}(x) = \frac{1}{\tau} \int_0^\infty dt\,  S_r(t)\, p(x,t),
\end{equation}
where \(\tau^{-1}\) ensures correct normalization.  
Specific to exponential \(r(t)\), we can use the relation \(S_r(t) = \tau\, r(t)\), which simplifies the above expression, 
\begin{equation}
n_{\text{off}}(x) 
= \int_0^{\infty} dt\, r(t)\, p(x,t) 
= \frac{1}{\sqrt{D\tau}}\, e^{ -|x| / \sqrt{D\tau} },
\label{eq:Laplace}
\end{equation}
which evaluates to a well-established Laplace distirbution ~\cite{PRL-Evans-2011,FP-Gupta-2022,JPA-Nagar-2023}.

To characterize return trajectories, we start with the propagator for a particle in a harmonic potential:
\begin{equation}
G(x, x_0, t) 
= \sqrt{\frac{ \mu K }{2\pi D \bigl( 1 - e^{-2 \mu K t} \bigr) }} \,
\exp\!\left[ -\frac{ \mu K \bigl( x - x_0 e^{-\mu K t} \bigr)^2 }
                   { 2D \bigl( 1 - e^{-2 \mu K t} \bigr) } \right],
\label{eq:G}
\end{equation}
which is the solution of 
\(
\dot G(x, x_0, t) =  \left[ \mu K x\, G(x, x_0, t) \right]' + D G''(x, x_0, t),
\)
subject to the initial condition \( G(x, x_0, 0) = \delta(x - x_0) \). 

Like \(p(x,t)\), the propagator \(G(x,x_0,t)\) does not capture the gradual disappearance 
of particles. In the return stage, particles vanish upon first arrival at the origin, 
which is implemented via an absorbing boundary at \(x=0\).  We incorporate such an adsorbing walls using 
the method of images, 
\[
G_a(x,x_0,t) = G(x,x_0,t) - G(x,-x_0,t).
\]
The propagator \(G_a\) is not normalized, and
\(
S(t|x_0) = \int_0^\infty dx\, G_a(x,x_0,t)
\)
is the survival probability of a particle starting at \(x_0\).

Because return trajectories originate from positions distributed according to \(n_{\text{off}}(x_0)\), 
the steady-state distribution is
\begin{equation}
n_{\text{on}}(x) = \int_0^\infty dx_0\, n_{\text{off}}(x_0) 
\left[ \frac{1}{\tau_{\text{on}}} \int_0^\infty dt\, G_a(x,x_0,t) \right],
\label{eq:n-on-integral}
\end{equation}
where \(\tau_{\text{on}}\) is the mean return time (or $1/\tau_{\text{on}}$ is the rate with which particles are adsorbed).  
Note that in the above expression we define $n_{\text{on}}$ on the positive half-space $x\geq 0$, which is sufficient 
due to the even symmetry $n_{\text{on}}(x) = n_{\text{on}}(-x)$.  
The distribution in Eq. (\ref{eq:n-on-integral}) is the solution of the following differential equation:
\begin{equation}
0 = [ \mu K x n_{\text{on}} ]' + D n_{\text{on}}'' + \frac{1}{\tau_{\text{on}}} n_{\text{off}}(x),
\label{eq:PDE-non}
\end{equation}
with absorbing boundary condition at \(x = 0\).  
Evaluating Eq.~\eqref{eq:n-on-integral} gives:
\begin{align}
n_{\text{on}}(x) &= 
   \frac{1}{2}\,  \sqrt{\frac{\pi}{2}}\,  \sqrt{\frac{1}{D \tau_K}} \, \frac{\tau_K}{\tau_{\text{on}}}  \, \exp\left[ -\frac{1}{2} \frac{\tau_K}{\tau} \right] \, 
   \exp\left[ - \frac{1}{2} \frac{x^2}{D\tau_K} \right]  \nonumber\\
&\quad \times \left[ \operatorname{erfi}\!\left(\sqrt{\frac{1}{2} \frac{\tau_K}{\tau}}\right) 
          - \operatorname{erfi}\!\left(\sqrt{\frac{1}{2} \frac{\tau_K}{\tau}} 
             - \sqrt{ \frac{1}{2} \frac{x^2}{D\tau_K}} \right) \right].
\label{eq:rho-on}
\end{align}
Here, \(\tau_K = 1/(\mu K)\) is the relaxation time in the harmonic trap.
Normalization \(\int_{0}^\infty dx\, n_{\text{on}}(x) = 1\) allows us to calculate $\tau_{\text{on}}$, 
\begin{equation}
\frac{\tau_{\text{on}}}{\tau_K}
= \frac{1}{2}\,e^{-\alpha/2}
\left[
\pi\,\operatorname{erfi}\!\sqrt{\frac{\alpha}{2}}
- \operatorname{Ei}\!\left(\frac{\alpha}{2}\right)
\right],
\label{eq:tau-on-0}
\end{equation}
where we introduced a dimensionless rate at which outgoing trajectories are terminated, 
\[
\alpha = \frac{\tau_K}{\tau}.
\]
The result in Eq. (\ref{eq:rho-on}) agrees with the derivation based on Fokker–Planck formalism in ~\cite{JOPA-Gupta-2020}, 
confirming the accuracy of the integral equation formulation in Eq. (\ref{eq:n-on-integral}) based on $G_a$.

Once \(\tau_{\text{on}}\) is known, the fraction of particles in the “off” state becomes:
\begin{equation}
f = \frac{\tau}{\tau + \tau_{\text{on}}},
\label{eq:f-off-on}
\end{equation}
and the full steady-state distribution reads:
\[
n(x) = f\, n_{\text{off}}(x) + (1-f)\, n_{\text{on}}(x).
\]

\begin{figure}[hhhh]
\centering
\begin{tabular}{rrrr}
\includegraphics[height=0.19\textwidth,width=0.23\textwidth]{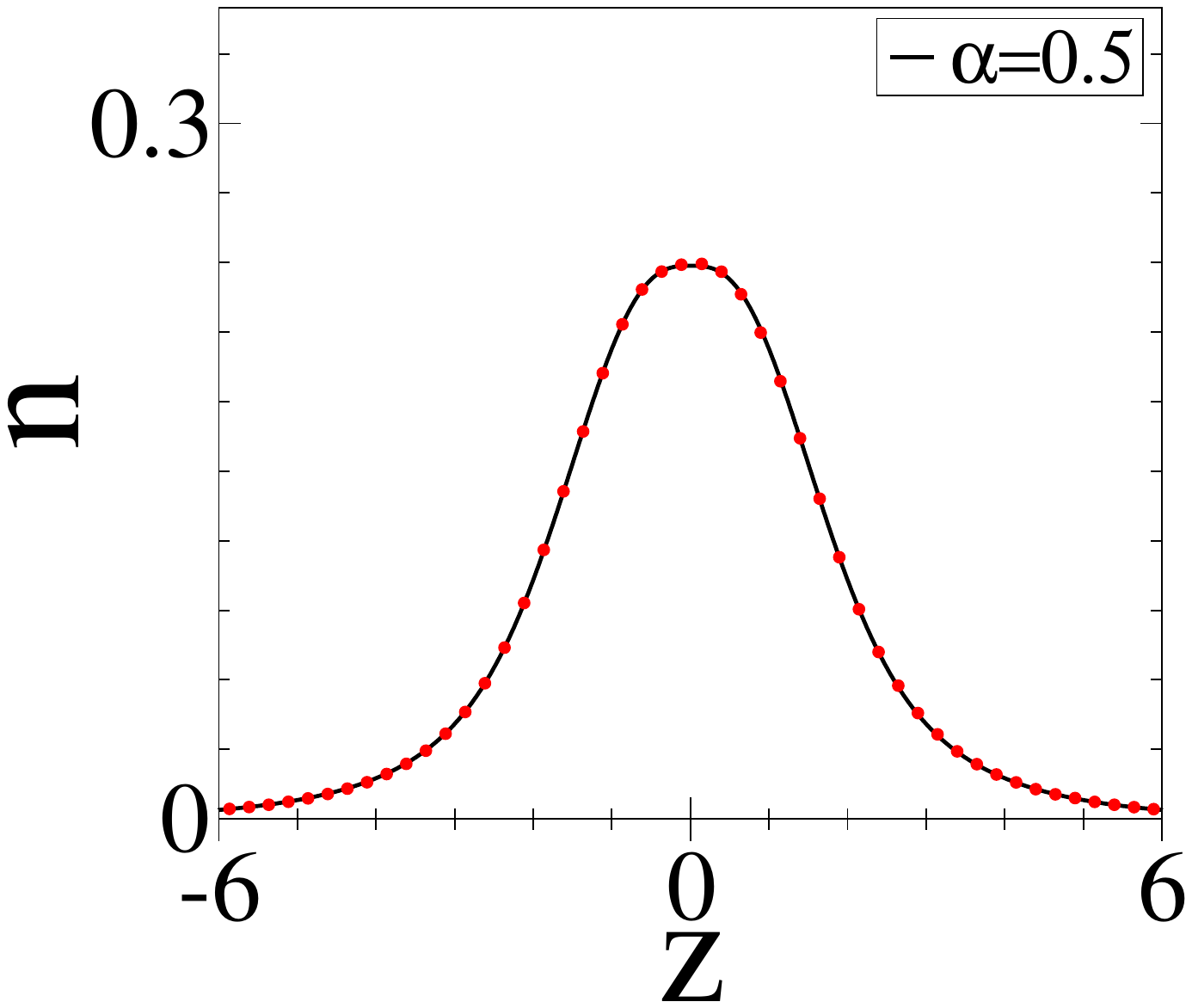} &&
\hspace{-0.25cm}\includegraphics[height=0.19\textwidth,width=0.23\textwidth]{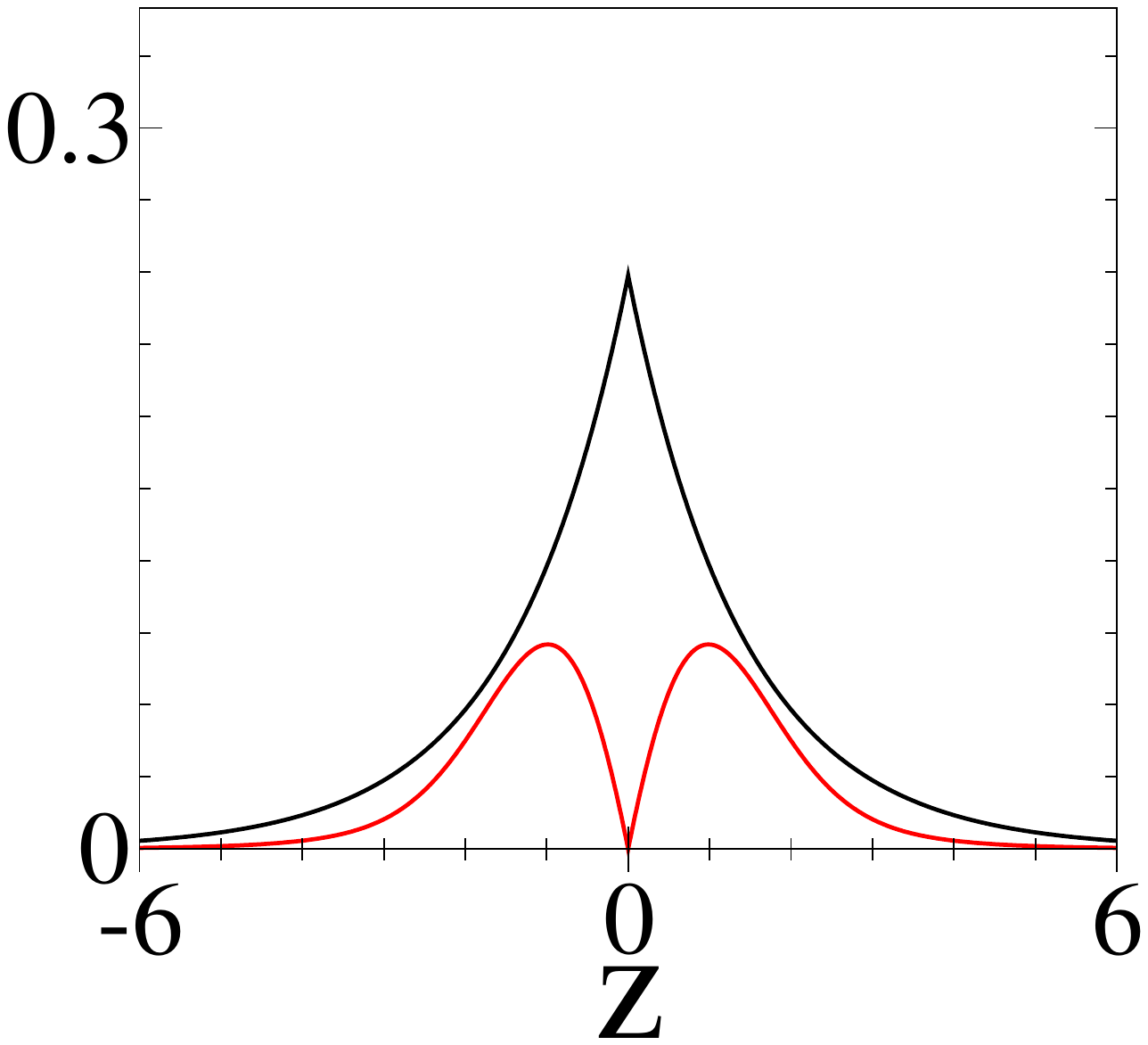} \\
\includegraphics[height=0.19\textwidth,width=0.23\textwidth]{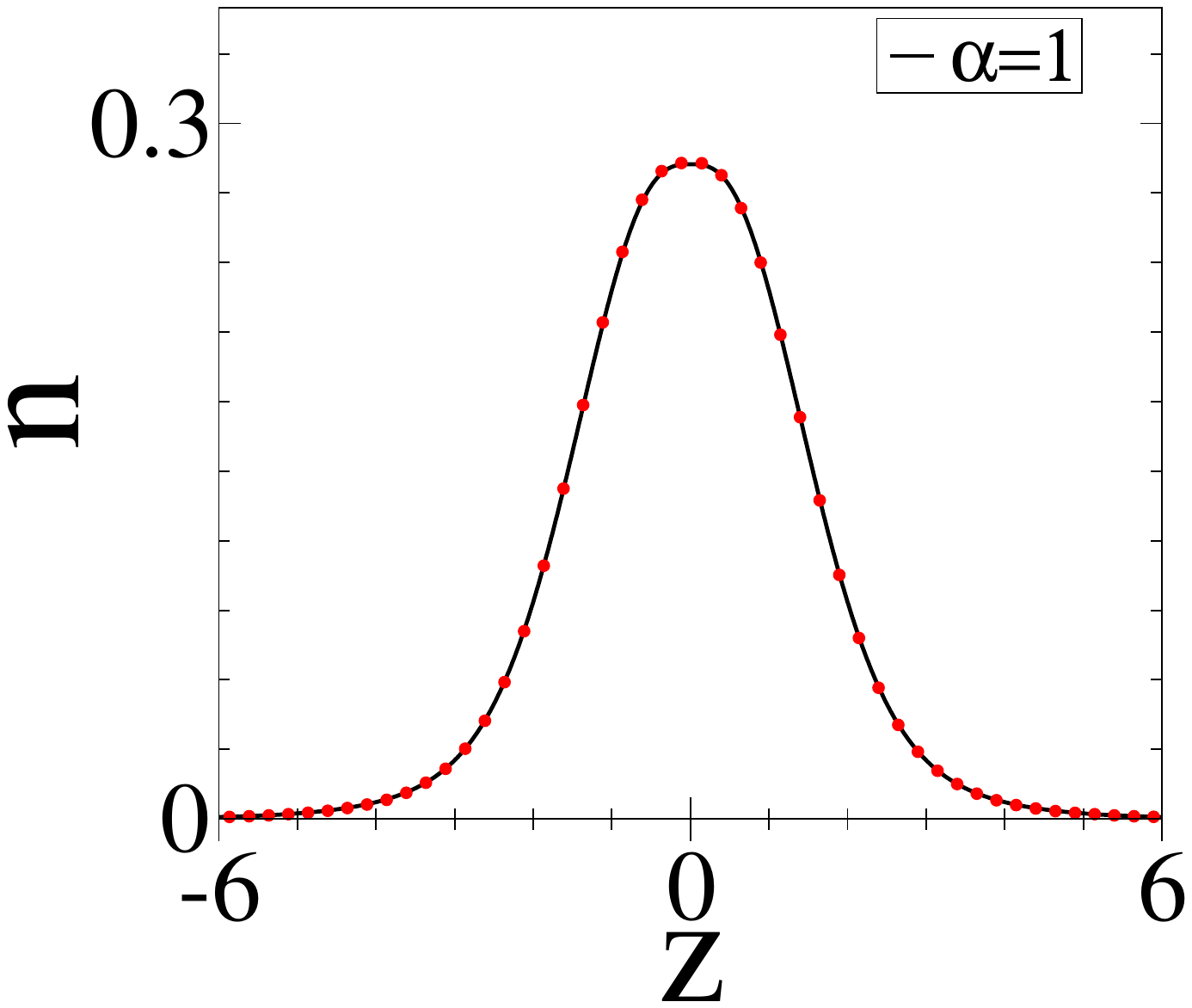} &&
\hspace{-0.25cm}\includegraphics[height=0.19\textwidth,width=0.23\textwidth]{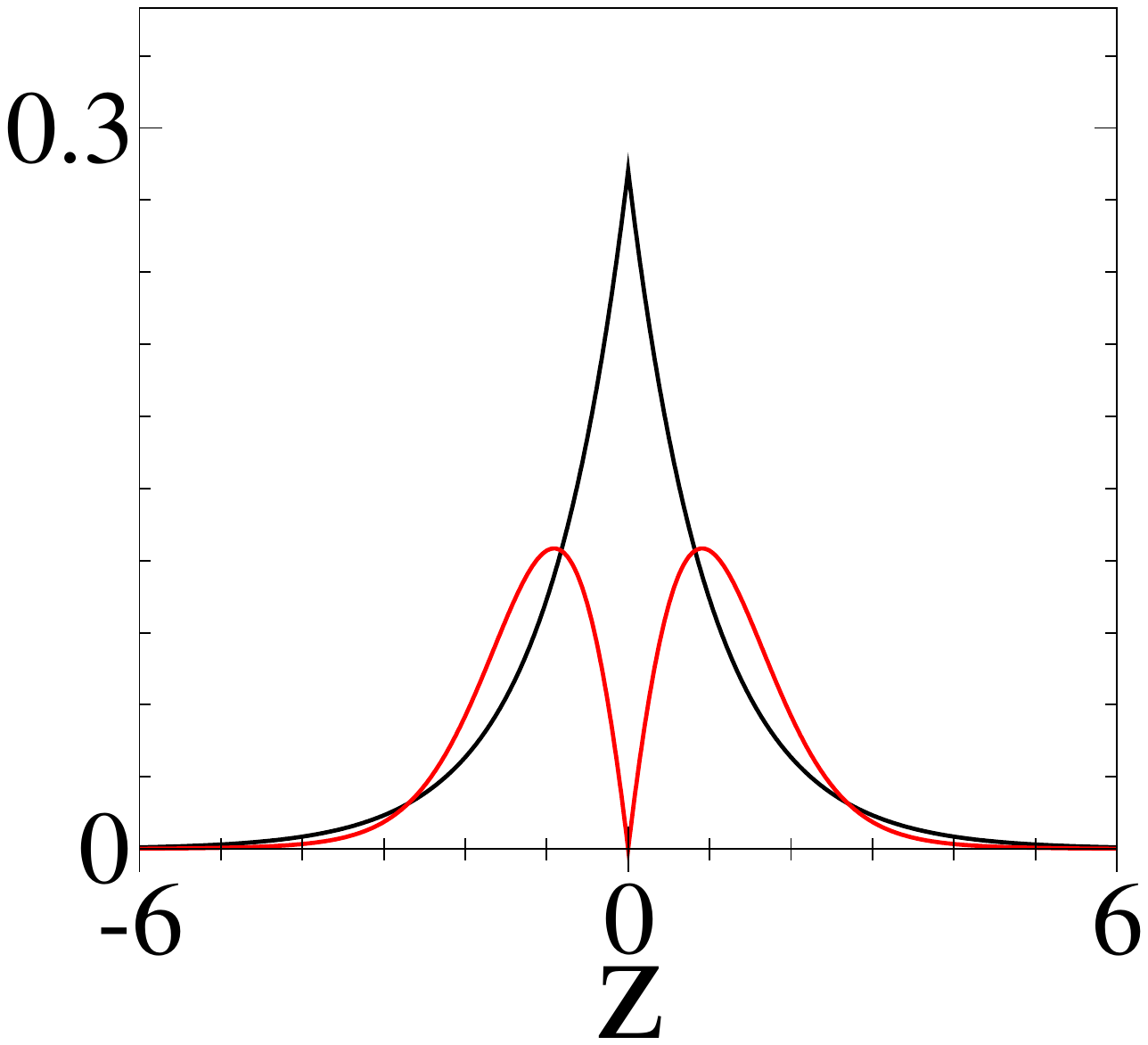} \\
\includegraphics[height=0.19\textwidth,width=0.23\textwidth]{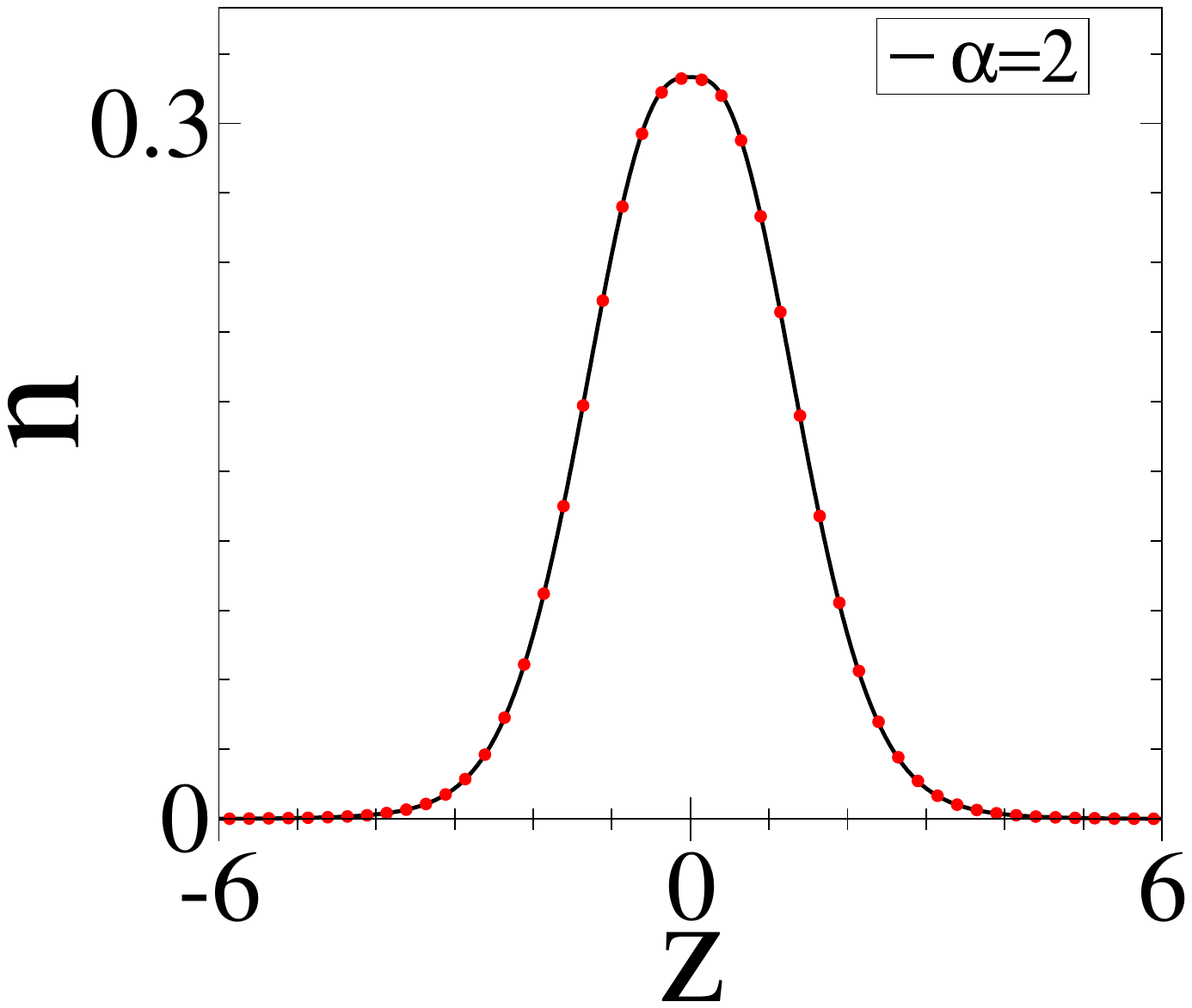} &&
\hspace{-0.25cm}\includegraphics[height=0.19\textwidth,width=0.23\textwidth]{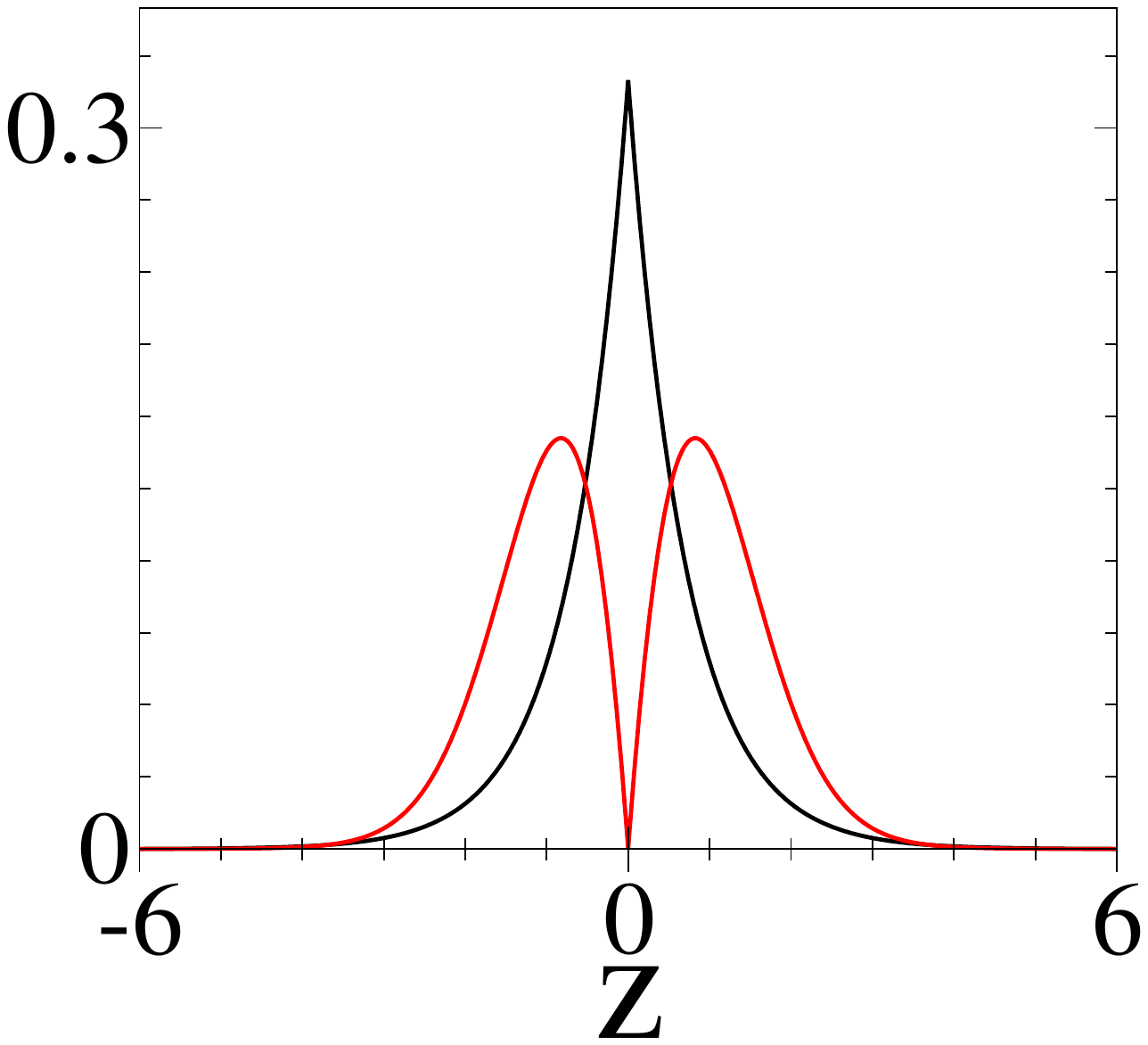}
\end{tabular}
\caption{
Steady-state distributions as a function of dimensionless distance \(z = x/\sqrt{D\tau_K}\) for various values of \(\alpha\). 
The left column shows the contributions from the two states: black lines correspond to the "off" state distribution 
\(f\, n_{\text{off}}\) as given in Eq.~(\ref{eq:Laplace}), and red lines correspond to the "on" state distribution 
\((1-f)\, n_{\text{on}}\) given in Eq.~(\ref{eq:rho-on}). 
The right column shows the total normalized distribution 
\(n = f\, n_{\text{off}} + (1-f)\, n_{\text{on}}\), where red circles denote simulation data, confirming analytical predictions.
}
\label{fig:rhoB}
\end{figure}

\begin{figure}[hhhh] 
 \begin{center}
 \begin{tabular}{rrrr}
\includegraphics[height=0.19\textwidth,width=0.23\textwidth]{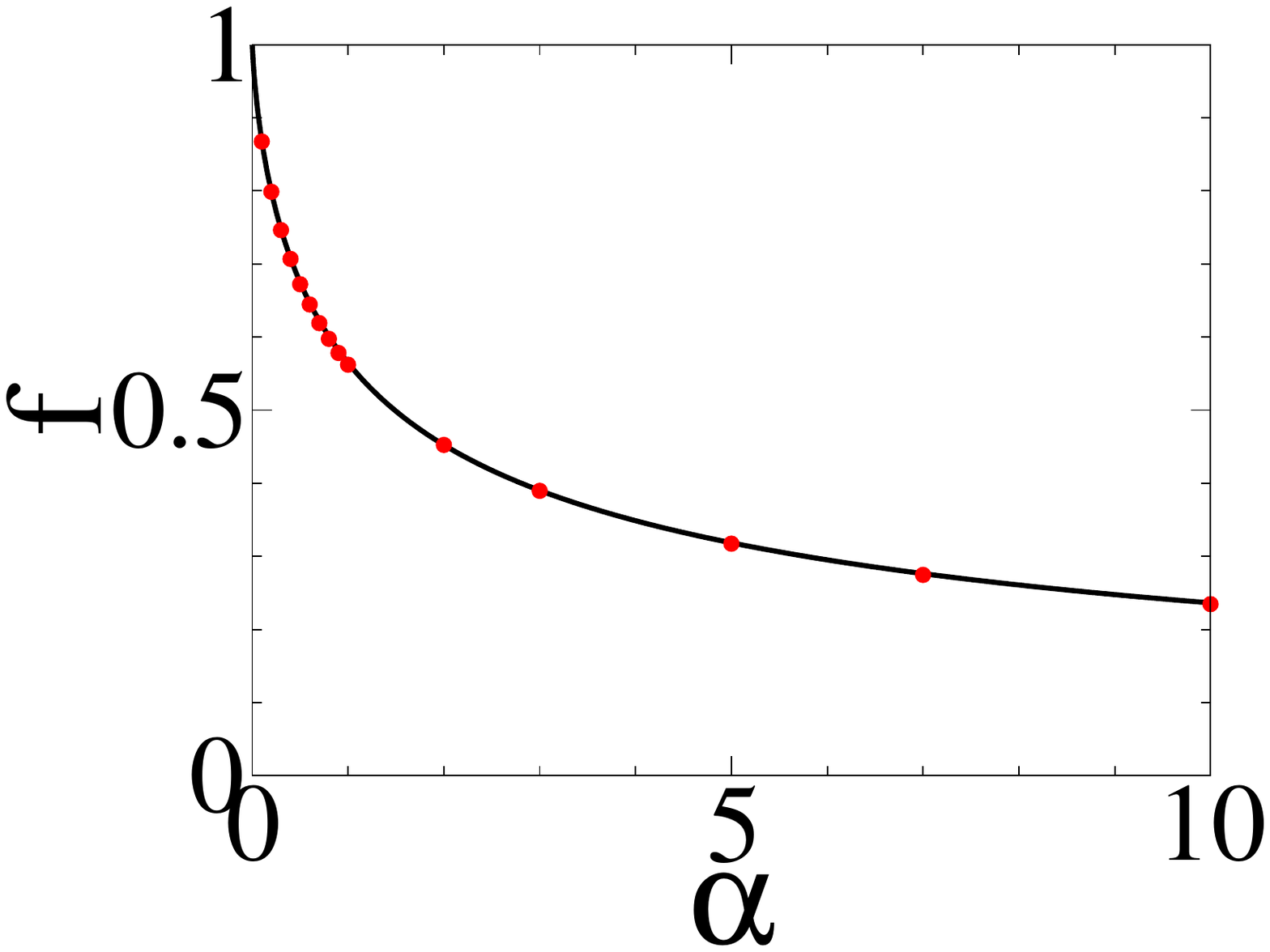} 
 \end{tabular}
 \end{center} 
\caption{ Fraction of particles in the ``off'' state, $f$, as a function of $\alpha = \tau_K/\tau$, where $\tau_K = 1/(\mu K)$.  
Red circles are simulation results.}
\label{fig:foff} 
\end{figure}

Both \(n_{\text{off}}\) and \(n_{\text{on}}\) exhibit a cusp at \(x = 0\). The two cusps cancel each other when combined, 
producing a flat top in $n(x)$, as seen in the left panels of Fig.~(\ref{fig:rhoB}).  
The probability of finding a particle in the "off" state is plotted in Fig.~(\ref{fig:foff}) as a function of 
a dimensionless rate of renewal \(\alpha = \tau_K/\tau\).  
At a slow rate, \(\alpha \to 0\), corresponding to long duration times of outgoing trajectories, \(f\) approaches $1$
as most particles are found in the ``off'' state.  
Then, as the rate \(\alpha\) increases, \(f\) slowly decays to zero as
\(
f \sim \sqrt{{2}/{\alpha \pi}}. 
\)
\graphicspath{{figures/}}

We note that the absorbing-boundary propagator of the OU process, $G_{a}(x,x_0,t)$, is restricted to the boundary at $x=0$.  
For an adsorbing boundary placed away from the origin, the source and image evolve with different velocities, producing a moving 
boundary.  Attempts to correct this with a weight factor do not work, since such a factor is time-dependent, 
and the resulting image term would no longer solve the same Fokker--Planck equation.  
In those cases, one must resort to Laplace--transform or spectral--expansion techniques \cite{PR-Pal-2020}.

Finally, we draw attention to an interesting property of the distribution \(\rho_{\text{on}} = (1 - f)\, n_{\text{on}}\).  
While \(\rho_{\text{on}}\) is not normalized—the integral 
\(\int_{-\infty}^{\infty} dx\, \rho_{\text{on}}(x) = 1 - f\) gives the probability of finding the particle in the ``on'' state—
its second moment, when calculated in dimensionless coordinates \(z = x/\sqrt{D \tau_K}\), is normalized:
\begin{equation}
\int_{-\infty}^{\infty} dz\, z^2\, \rho_{\text{on}}(z) = 1.
\end{equation}
This identity can be directly verified by applying the operator \(\int dx\, x^2\) to Eq.~(\ref{eq:PDE-non}).

A direct implication is that the total average potential energy of the system is the same as that of a particle 
in thermal equilibrium:
\begin{equation}
PE = \int_{-\infty}^{\infty} dx\, \frac{Kx^2}{2}\, \rho_{\text{on}}(x) = \frac{k_B T}{2},
\label{eq:PE}
\end{equation}
Particles in the ``off'' state do not contribute to the potential energy since the trap is turnd off during that phase.

This result holds for any $r(t)$, provided the trap is harmonic.  It is notable because it connects 
an equilibrium property to a non-equilibrium steady state.  

%

\subsection{waiting time distributions}
\label{sec:sec3}

Other quantities of interest are the waiting time distributions.  Using the propagator \(G_a(x,x_0,t)\), 
we define the survival function 
\(
S(t|x_0) = \int_{0}^{\infty}dx\, G_a(x,x_0,t),
\)
related to the first-passage-time distribution as \(R_{\text{on}} = -\dot{S}\), and leading to
\be
R_{\text{on}}(s|z_0) =  e^{2s} \sqrt{\frac{ z_0^2}  {4 \pi  \left(e^{s} \sinh s\right)^3}}  
\exp \left(-\frac{ z_0^2 }{4 e^{s} \sinh s }\right), 
\label{eq:R-on}
\ee
where we use the dimemnsionless variables of distance $z = x/\sqrt{D\tau_K}$ and time $u = t/\tau_K$, 
where $\tau_K = 1/(\mu K)$.
\(R_{\text{on}}(s|z_0)\) represents the distribution of waiting times it takes a particle, initially at \(z_0 = x_0/\sqrt{D\tau_K}\), 
to reach the origin for the first time.  

From \(R_{\text{on}}(s|z_0)\) we  calculate the mean time, 
\[
\langle s\rangle_{z_0} =  \int_0^{\infty} ds\, s\, R_{\text{on}}(s|z_0),
\]
which, after evaluation, becomes
\begin{equation}
\langle s\rangle_{z_0} 
 = \frac{1}{2} \left[
\pi\,\operatorname{erfi}\!\left(\frac{z_0}{\sqrt{2}}\right)
- z_0^{2}\,
{}_2F_2\!\left(1,1;\tfrac{3}{2},2;\tfrac{z_0^{2}}{2}\right)
\right].
\label{eq:s-z0}
\end{equation}
Since this quantity will be frequently used in defining the mean first passage time of various protocols, 
below we provide an explicit expression in physical parameters, 
\begin{equation}
t_{\text{on}}(x_0)
 = \frac{\tau_K}{2} \left[
\pi\,\operatorname{erfi}\!    \sqrt{\frac{x_0^{2}}{2D\tau_K}} 
- \frac{x_0^{2}}{D\tau_K} \,
{}_2F_2\!\left(1,1;\tfrac{3}{2},2; \tfrac{x_0^{2}}{2D\tau_K} \right)
\right], 
\label{eq:s-x0}
\end{equation}
where $t_{\text{on}}$ is related to $ \langle s\rangle_{z_0}$ via $t_{\text{on}} = \langle s\rangle_{z_0} \tau_K$.  
The above expression represents the mean time to reach $x=0$ for the first time for a particle in a harmonic
trap and initially at $x_0$.

The waiting time distribution that a particle remains in the "on" state, designated by \(r_{\text{on}}(u)\), 
is obtained by averaging \(R_{\text{on}}\) over the distribution of initial positions:
\begin{equation}
r_{\text{on}}(u) = \int_{-\infty}^{\infty} dz_0\, n_{\text{off}}(z_0)\, R_{\text{on}}(s|z_0).
\label{eq:r-on-gen}
\end{equation}
Inserting \(n_{\text{off}}(z)\) and \(R_{\text{on}}(s|z_0)\) into Eq.~\eqref{eq:r-on-gen} yields
\begin{equation}
r_{\text{on}}(u) = \alpha\, e^{2s} 
\left[
\frac{1}{\sqrt{\pi\, \alpha\, e^s \sinh s}} 
- e^{\alpha\, e^s \sinh s} \, \mathrm{erfc}\!\left(\sqrt{ \alpha\, e^s \sinh s}\right)
\right], 
\label{eq:app1F}
\end{equation}
where $\alpha = \tau_K/\tau$.

\begin{figure}[h!] 
 \begin{center}
 \begin{tabular}{cc}
\includegraphics[height=0.19\textwidth,width=0.23\textwidth]{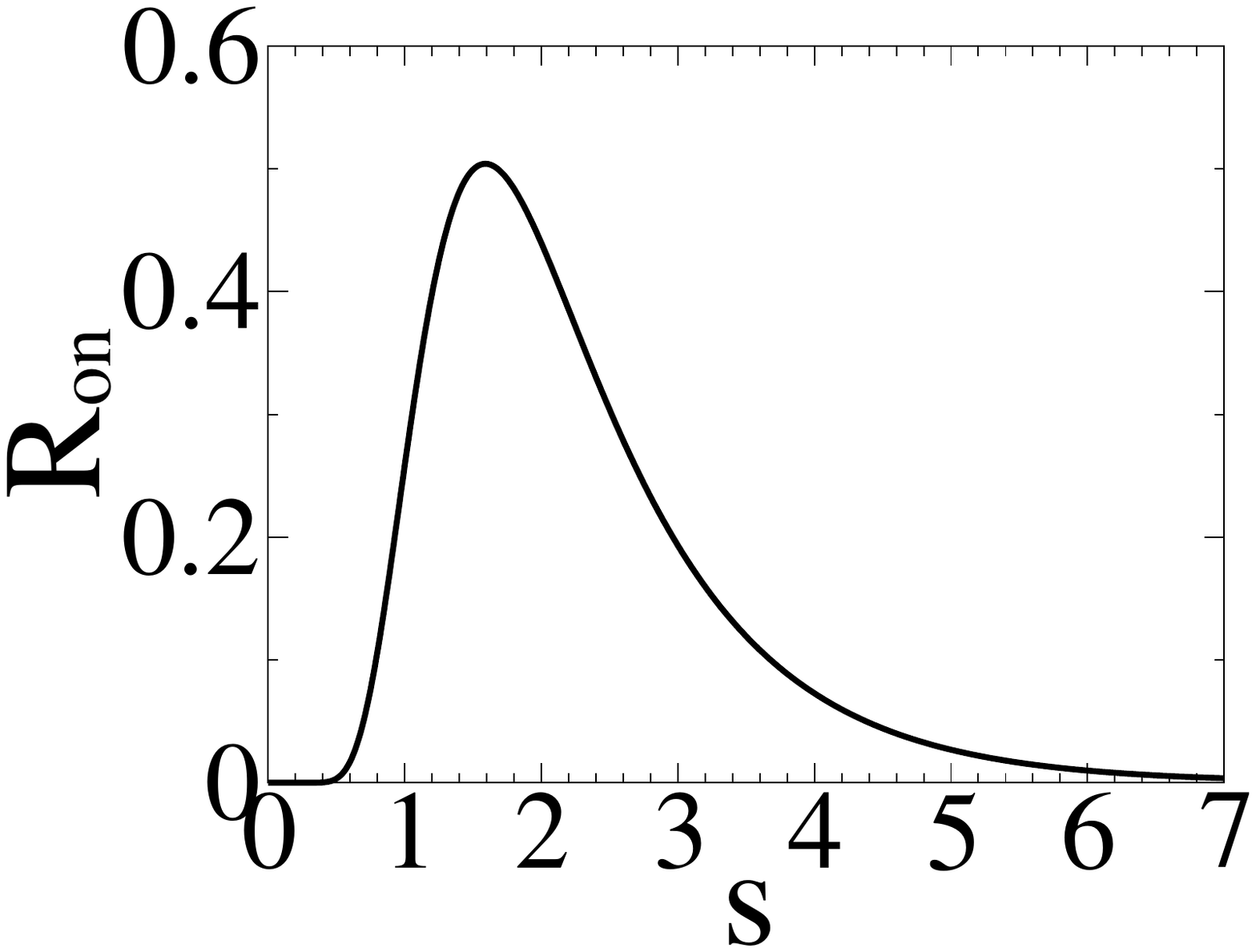}  &
\includegraphics[height=0.19\textwidth,width=0.23\textwidth]{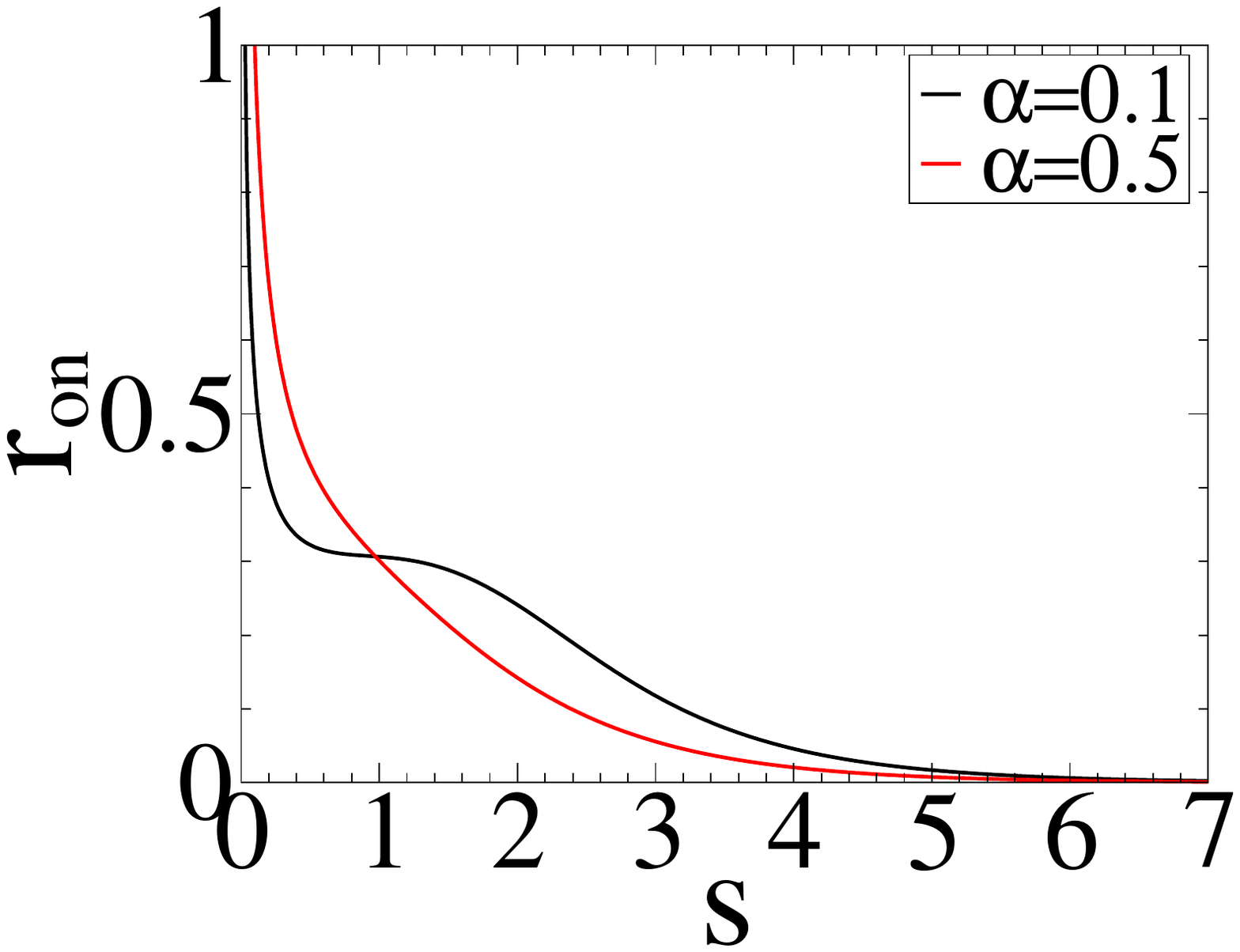}  
 \end{tabular}
 \end{center} 
\caption{
First-passage-time distributions as a function of dimensionless time \(u = t/\tau_K\). 
The left panel shows \(R_{\text{on}}(s\,|\,z_0)\) for a particle initially located at \(z_0 = 5\), 
while the right panel displays the total return-time distribution \(r_{\text{on}}(u)\) 
for different values of \(\alpha\). 
The exact expressions are given in Eq.~(\ref{eq:R-on}) for \(R_{\text{on}}(s,z_0)\) and 
in Eq.~(\ref{eq:app1F}) for \(r_{\text{on}}(u)\).
}
\label{fig:pL1} 
\end{figure}

In Fig.~(\ref{fig:pL1}) we plot the first-passage-time distributions \(R_{\text{on}}(s\,|\,z_0)\) and \(r_{\text{on}}(u)\).  
Although \(r_{\text{on}}\) is derived from \(R_{\text{on}}\), their behaviors at short times differ drastically.  

\graphicspath{{figures/}}
These features are evident from the respective asymptotic behaviors:
\begin{equation}
R_{\text{on}}(s,z_0) \sim 
\begin{cases}
\displaystyle \frac{e^{-z_0^2/4s}}{\sqrt{4\pi s^3}}\, z_0\, e^{z_0^2/4}, & \text{as } s \to 0, \\[10pt]
\displaystyle e^{-s} \, z_0 \sqrt{\frac{2}{\pi}}, & \text{as } s \to \infty,
\end{cases}
\label{eq:p-asym}
\end{equation}
and
\[
r_{\text{on}}(u) \sim 
\begin{cases}
\displaystyle \sqrt{\tfrac{\alpha}{\pi s}}, & \text{as } s \to 0, \\[8pt]
\displaystyle e^{-s} \sqrt{\tfrac{2}{\alpha \pi}}, & \text{as } s \to \infty.
\end{cases}
\]
While both functions exhibit exponential decay at large \(s\), only \(r_{\text{on}}(u)\) diverges for small \(s\). 
This divergence originates from the structure of the distribution \(n_{\text{off}}\), 
which allows initial positions arbitrarily close to \(z_0 = 0\), yielding return times near zero.

\subsection{Time--Energy Trade-Off}
\label{sec:sec4}

In this section, we briefly look at the energy consumption of return trajectories due to the external driving force.  

The class of models in which return trajectories are terminated at random times drawn from a prescribed distribution 
$r_{\text{on}}(t)$---in contrast to our protocol, where $r_{\text{on}}(t)$ emerges self-consistently---has two main drawbacks.  
First, the outgoing trajectories originate from a broad distribution of initial positions rather than from the fixed point $x=0$.  
Second, the randomly sampled return times are often too short for this distribution to relax to equilibrium;  
consequently, the initial-position distribution is not Gaussian but dominated by exponential tails extending far from the origin.  

One can partially correct this issue by allowing the return trajectories sufficient time to equilibrate, 
so that $n_{\text{on}}$ approaches a Boltzmann distribution, $n_{\text{on}} \propto e^{-x^2/2D\tau_K}$ \cite{PRR-Besga-2020}.  
Although this does not ensure that the outgoing trajectories start exactly from $x=0$,  
it provides a substantial improvement over a distribution that has an exponential tail.  
From Eq.~(\ref{eq:G}), we know that the mean position of a particle in a harmonic trap decays as  
$\langle x(t)\rangle = x_0 e^{-t/\tau_K}$.  
A common rule of thumb based on this result is that equilibration is effectively reached at  
$\tau_{\mathrm{eq}}\!\approx\!3\tau_K$.

\begin{figure}[hhhh] 
 \begin{tabular}{rrrr}
\includegraphics[height=0.23\textwidth,width=0.27\textwidth]{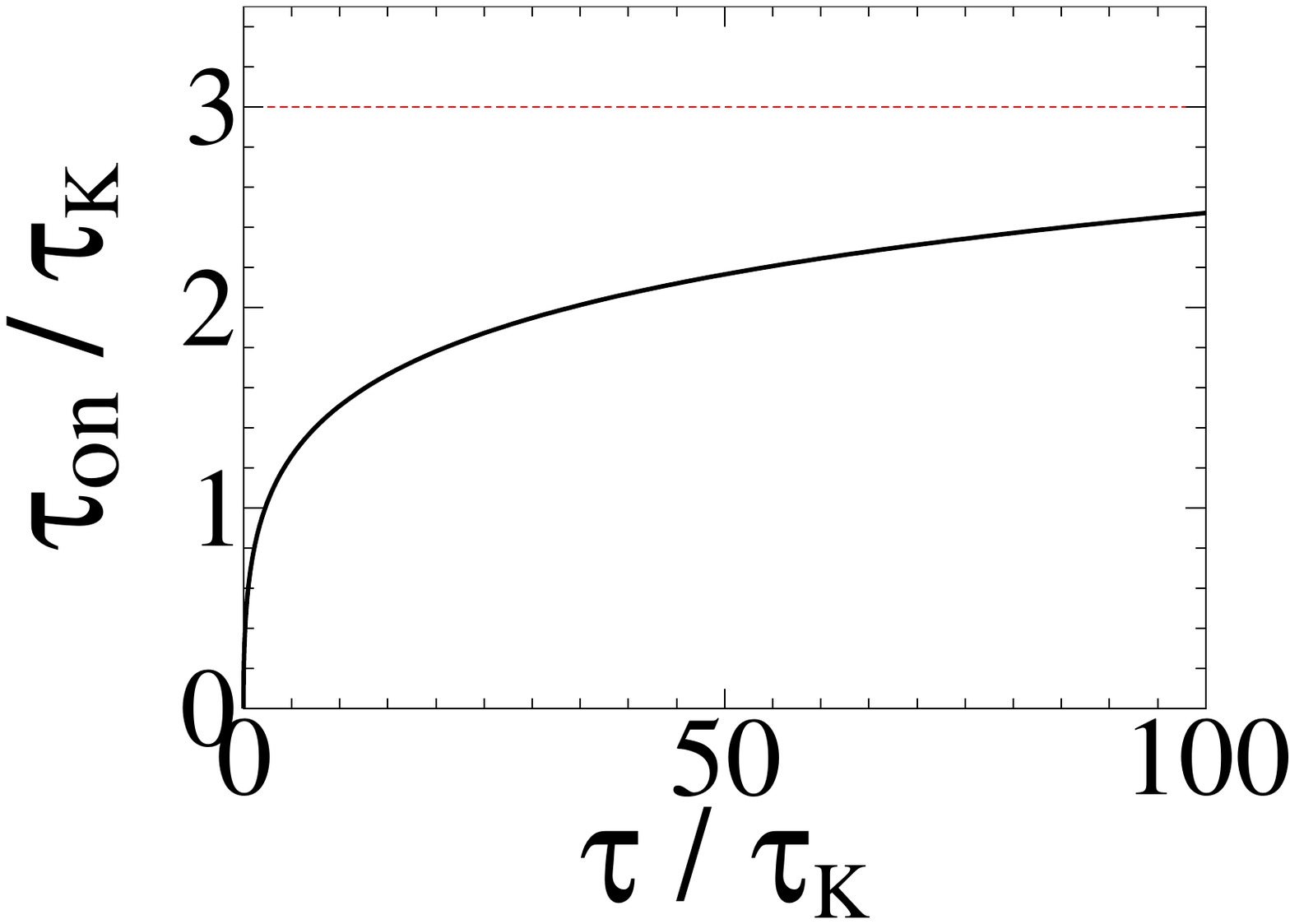} 
 \end{tabular}
\caption{Mean return time $\tau_{\text{on}}$ as a function of $\tau$.  
The dashed line marks the passive equilibration time $\tau_{\mathrm{eq}}\!\approx\!3\tau_K$.
$\tau_{\text{on}}$ remains below $\tau_{\mathrm{eq}}$ except for very large ratios $\tau/\tau_K$.}
\label{fig:ton-tau} 
\end{figure}

To see how $\tau_{\text{on}}$ of our model compares with $\tau_{\mathrm{eq}}$, in Fig.~(\ref{fig:ton-tau}) we plot $\tau_{\text{on}}$ 
as a function of $\tau$, 
We observe that $\tau_{\text{on}}$ remains below $\tau_{\text{eq}}$, 
except for very large values of $\tau$, approximately $\tau \gtrsim 300\,\tau_K$.  
Thus, our protocol, in addition to ensuring the precise starting point of outgoing trajectories,  
is faster compared to passive equilibration.  
\graphicspath{{figures/}}

To understand the behavior of $\tau_{\text{on}}$ at large $\tau/\tau_K$ (corresponding to large $K$), 
we examine its asymptotic form in the limit of large $K$:
\begin{equation}
\frac{\tau_{\text{on}}}{\tau_K} \sim \frac{1}{2}\,\ln\!\frac{\tau}{\tau_K}.
\label{eq:TK}
\end{equation}
The fact that $\tau_{\text{on}}$ eventually exceeds $\tau_{\mathrm{eq}}$ follows from the logarithmic factor
in Eq.~\eqref{eq:TK}, which is a hallmark of first-passage in a harmonic trap. 
Its origin can be traced to the mean first-passage time conditioned on the initial offset $z_0=x_0/\sqrt{D\tau_K}$:
\[
\langle s \rangle_{z_0}
= \int_0^\infty s\,R_{\text{on}}(s\,|\,z_0)\,ds
\sim \ln z_0 + \tfrac{1}{2}\bigl(\gamma + \ln 2\bigr),
\qquad (z_0\gg 1),
\]
so that averaging $\langle s\rangle_{z_0}$ over the initial-position
distribution $n_{\text{off}}(x_0)$ produces the overall logarithm in Eq.~\eqref{eq:TK}.
The absence of a logarithmic term in $\tau_{\mathrm{eq}}$ has more to do with the convention 
$\tau_{\mathrm{eq}} \approx 3\tau_K$.  A more accurate definition of $\tau_{\mathrm{eq}}$ would likely  
yield a similar logarithmic factor.  However, we will not pursue refinements of the definition of $\tau_{\mathrm{eq}}$ in this work
and use $\tau_{\mathrm{eq}} \approx 3\tau_K$ as sufficient for what we wish to demonstrate.

We next turn to the question of energy.  The parameter $K$ gives us control over how fast the return process
is complete.  
By increasing $K$, $\tau_{\text{on}}$ can be reduced to an arbitrarily low value.  
The key question, however, is: at what energetic cost?  

The instantaneous energy of a particle when the trap is switched on is 
\(
u = \frac{1}{2} K x_0^{2},
\)
where $x_0$ is the initial position of the return trajectory.  
The trajectory terminates when the particle reaches the origin, at which point this energy has been entirely dissipated as heat.  
Averaging over the distribution of initial positions $n_{\text{off}}(x_0)$ (see Eq.~\ref{eq:r-on-gen}),  
the mean dissipated energy per cycle is
\be
 \Delta E = \int_{-\infty}^{\infty} dx_0\, n_{\text{off}}(x_0)\, u(x_0) = D K \tau.  
\label{eq:u-ave}
\ee

We now compare this result with the energy balance for particles equilibrating inside a harmonic potential.  
Since outgoing trajectories in this case start from a Boltzmann distribution, 
the particles are more spatially spread at the moment the trap is switched on.  
The mean potential energy at that instant is 
\(
\langle u(x_0)\rangle_1 = D K \left(\tau + \tfrac{\tau_K}{2}\right),
\)
where the second term accounts for the fact that particles do not start from the origin but from an equilibrium distribution.  
At the end of equilibration, when the potential is switched off, the equilibrium mean energy is 
\(
\langle u(x_0)\rangle_2 = D K \tfrac{\tau_K}{2}.
\)
Taking the difference, $\Delta E = \langle u(x_0)\rangle_1 - \langle u(x_0)\rangle_2$, 
recovers the result in Eq.~\eqref{eq:u-ave}.  
Hence, our first-passage-based protocol dissipates the same amount of heat as the protocol 
based on full equilibration.  The advantage is that our protocol does it more precisely, 
by bringing particles to the origin, and doing it faster.

The above analysis suggests a speedup relative to passive equilibration 
without any additional mechanical energy cost.  
It is important to mention another protocol, the Engineered Swift Equilibration (ESE) scheme 
proposed in~\cite{Nature-Trizac-2016,PRR-Besga-2020}.  
In ESE, the system is driven to equilibrium within a prescribed time 
by engineering a time-dependent potential $K(t)$.  
While ESE achieves fast equilibration, it necessarily incurs a higher energetic cost.  
In contrast, our protocol attains a comparable speedup without mechanical energy penalties.

While ESE accelerates relaxation by injecting energy,
our protocol uses \emph{information} to accelerate the return phase.  
This places our model in the broader class of Maxwell-demon systems such as 
the information ratchet ~\cite{PRL-Sagawa-2010,PRL-Jarzynski-2025} 
or the Szilard’s engine ~\cite{Parrondo2015}.

\subsubsection{entropy production}

Having determined the amount of energy injected during a single cycle, the steady-state rate of heat dissipation, 
$\dot q$, can be obtained by dividing $\Delta E$ by the mean duration of a cycle, $\tau + \tau_{\text{on}}$. 
Using the standard thermodynamic relation $\dot q = T\Pi$, where $\Pi$ is the entropy production rate and $T$ is the temperature, we find:
\begin{equation}
T \Pi = \dot q = \frac{\Delta E}{\tau + \tau_{\text{on}}} = D K \left( \frac{\tau}{\tau + \tau_{\text{on}}} \right) = D K f,
\label{eq:Pi}
\end{equation}
where $f$ denotes the fraction of particles in the “off” state under steady-state conditions.
Somewhat counterintuitively, $\Pi$ increases as the particle spends more time in the “off” state. 


The entropy production of stochastic resetting processes in arbitrary external potentials has been analyzed from a formal perspective 
in Ref.~\cite{PRR-Mori-2023}. Our derivation is more intuitive and for a specific system.  Other formal studies related to the entropy 
production rate or other non-equilibrium aspects of stochastic resetting can be found in \cite{EPL-Fuchs-2016,PRE-Pal-2017,PRL-Pal-2020}.

\section{The mean first passage time}
\label{sec:sec5}

An important feature of the SR protocols is their ability to optimize search efficiency by eliminating 
unproductive trajectories. Rather than allowing a particle to wander indefinitely when the target is not found, the system resets: 
the particle is returned to its initial position, and a new search begins. This interruption mechanism is key to improving the mean time to locate a target.

In this section, we analyze four distinct SR protocols. The difference between them lies in how the return trajectories are used.  
These protocols are:
\begin{itemize}
    \item \textbf{SR:} The basic resetting protocol, with outgoing trajectories and without return trajectories.

    \item \textbf{SR-A:} A protocol that includes a reverse trajectory; however, the trajectory is \emph{invisible} to the target, so the target cannot be located during a return motion.

    \item \textbf{SR-B:} Similar to SR-A, but with a \emph{visible} return trajectory. As a result, the target can be located by the return trajectory.

    \item \textbf{SR-C:} A protocol involving only return trajectories. As soon as the particle reaches \( x = 0 \), it is instantly transferred to a new starting point \( x_0 \ne 0 \), 
    and the next return trajectory is initiated.
\end{itemize}

Our approach to computing the MFPT is based on decomposing the full resetting process into generations of 
trajectories and then decomposing each process within each generation into phenomenological 
parameters and characteristic timescales. Since the target is located at \(x = L\), and the return termination occurs at \(x = 0\), 
the process is naturally framed as a two-boundary problem. These two absorbing boundaries give 
rise to a splitting probability scenario. Each parameter is then defined and computed within this framework.

The idea of using return trajectories as an active part of the search process has recently been explored in~\cite{PRE-Arnab-2024,PRE-Arnab-2025}. 
Our approach differs both in derivation and in scope. Rather than analyzing a single protocol, we construct and 
compare several search protocols based on the same potential landscape. Furthermore, our proposal of the SR-C 
protocol introduces a novel perspective: a search mechanism that relies solely on return motion, bypassing the traditional 
outward exploration phase altogether.

\subsection{definitions and notation}

In this section we collect the parameters and distributions defined in the previous section. 
These quantities will be used to derive the MFPT for the various protocols discussed later in the paper.

\vspace{0.3cm}
\noindent\textbf{Time scales:}
\begin{list}{}{\leftmargin=0pt \itemindent=0pt \itemsep=4pt}
\item $\tau$ — average duration of an outgoing trajectory, corresponding to the exponential distribution 
      $r(t) = \tau^{-1} e^{-t/\tau}$.
\item $\tau_K = 1/(\mu K)$ — Ornstein--Uhlenbeck relaxation time of a harmonic trap of stiffness $K$.
\item $\tau_{\text{on}}$ — average duration of a return trajectory drawn from the distribution $r_{\text{on}}(t)$
and defined as 
\be
\frac{\tau_{\text{on}}}{\tau_K}
= \frac{1}{2}\,e^{-\alpha/2}
\left[
\pi\,\operatorname{erfi}\!\sqrt{\frac{\alpha}{2}}
- \operatorname{Ei}\!\left(\frac{\alpha}{2}\right)
\right],
\label{eq:tau-on-0-2}
\ee
where $\alpha = \tau_K/\tau$.  

\item $t_{\text{on}}(x_0)$ --- the mean first passage time of a particle in a harmonic trap initially in $x_0$, 
defined as 
\begin{equation}
t_{\text{on}}(x_0)
 = \frac{\tau_K}{2} \left[
\pi\,\operatorname{erfi}\!    \sqrt{\frac{x_0^{2}}{2D\tau_K}} 
- \frac{x_0^{2}}{D\tau_K} \,
{}_2F_2\!\left(1,1;\tfrac{3}{2},2; \tfrac{x_0^{2}}{2D\tau_K} \right)
\right].  
\label{eq:ton2}
\end{equation}

\end{list}

%
%

%
%

\vspace{0.25cm}\noindent 
The search process is partitioned into identical cycles, each composed of an outgoing 
trajectory (starting at $x=0$) and a return trajectory (ending at $x=0$).  
A particle may find the target during either stage.  
Table~\ref{tab:MFPT-quantities} summarizes the quantities used in defining the mean first passage time (MFPT).

\begin{table}[h!]
\centering
\begin{tabular}{l l}
\hline
$p$ & Probability that an outgoing trajectory is successful \\[3pt]
$1-p$ & Probability that an outgoing trajectory is unsuccessful \\[3pt]

$q$ & Probability that a return trajectory is successful \\[3pt]
$1-q$ & Probability that a return trajectory is unsuccessful \\[3pt]

$\tau_{p}$ & Mean duration of a successful outgoing trajectory \\[3pt]
$\tau_{1-p}$ & Mean duration of an unsuccessful outgoing trajectory \\[3pt]

$\tau_{q}$ & Mean duration of a successful return trajectory \\[3pt]
$\tau_{1-q}$ & Mean duration of an unsuccessful return trajectory \\[3pt]
\hline
\end{tabular}
\caption{Quantities characterizing the outgoing and return stages of each search cycle.}
\label{tab:MFPT-quantities}
\end{table}

\subsection{SR protocol}

We start with the basic SR protocol, which does not include return trajectories, and the moment 
that an outgoing trajectory is interrupted a particle appears at $x=0$.  
We revisit this protocol to introduce our approach for calculating the mean first-passage time (MFPT).  

Due to the cyclical nature of the SR protocol, it is useful to organize trajectories into \emph{generations}. 
Trajectories that have never been renewed are called \emph{first generation}; those renewed once are 
\emph{second generation}; and so on.  
The probability of finding a target by any single trajectory is $0 < p \leq 1$ and, therefore, the probability 
of missing a target is $1-p$.

If a targe is not found duing a first generation, it may be found at second, third, etc.  The probability that 
a target is found after $n$ attempts is defined as
\be
p_n = (1 - p)^{n - 1} p,
\label{eq:pn}
\ee
indicating that the last successful trajectory was preceded by $n-1$ unsuccessful trajectories.  
Summing over all possible generations, we get 
\begin{equation}
\sum_{n=1}^{\infty} p_n = 1,
\label{eq:S-pn}
\end{equation}
which shows that the protocol always finds the target as long as $0 < p \leq 1$.  

Next, we compute the mean time for a sequence of $n$ trajectories to find the target.  
Let $\tau_p$ denote the mean time of a single successful trajectory, and $\tau_{1-p}$ the mean time of a single unsuccessful 
trajectory.  Note that $\tau_{1-p} \neq \tau$ since it accounts only for unsuccessful trajectories.  
Considering that a target is found after $n$ attempts, the mean time due to all the $n$ outgoing trajectories is given by 
\begin{equation}
\tau_n = (n - 1)\,\tau_{1-p} + \tau_p,
\label{eq:Tn}
\end{equation}
where the $(n-1)$ unsuccessful trajectories each contribute $\tau_{1-p}$ and the final successful trajectory contributes $\tau_p$.

Having defined $p_n$ and $\tau_n$, the MFPT of the entire SR protocol can be written as
\begin{equation}
\tau_{\rm sr} = \sum_{n=1}^{\infty} p_n\,\tau_n,
\label{eq:tau_sr-gen}
\end{equation}
which evaluates to
\begin{equation}
\tau_{\rm sr} = \tau_p + \frac{1 - p}{p}\,\tau_{1-p},
\label{eq:tau_sr}
\end{equation}
which shows that $\tau_{\rm sr} \geq \tau_p$, which makes sense since on average the target is not 
found at a first attempt.  

Defining the average number of generations required to find the target as
\(
\langle n \rangle = \sum_{n=1}^{\infty} n\,p_n = p^{-1},
\)
the MFPT can be expressed as
\begin{equation}
\tau_{\rm sr} = \langle n \rangle \big[\, p\,\tau_p + (1 - p)\,\tau_{1-p} \big].
\end{equation}
Thus, the MFPT of the SR protocol is proportional to the average number of generations required to find a target 
multiplied by a weighted average of the two characteristic time scales.

\subsubsection{derivation of the MFPT expression} 

The MFPT for the SR-A protocol in Eq. (\ref{eq:tau_sr}) is defined in terms of three parameters: $p$, $\tau_{1-p}$, and $\tau_p$.  
We next show how these parameters emerge from underlying dynamics.

We start by defining the distribution of first-passage times for trajectories without stochastic interruptions, $h(t)$, which by being normalized, 
\(
\int_0^{\infty} dt\, h(t) \;=\;  1, 
\)
implies that eventually all trajectories find a target.  The second distribution needed is the distirbution of times 
a particle remains in an outgoing stage, $r(t)$, which for our particular model is an exponential distirbution.

Associated with both distributions $r(t)$ and $h(t)$ are the corresponding survival functions,
\begin{equation}
S_r(t) = \int_t^{\infty} dt'\, r(t'), 
\qquad
S_h(t) = \int_t^{\infty} dt'\, h(t').
\end{equation}

The probability $p$ of a successful trajectory and the corresponding mean time $\tau_p$ can then be defined in terms 
of those quantities as 
\begin{align}
p &= \int_0^{\infty} dt\, h(t)\, S_r(t), \nonumber\\[4pt]
p\,\tau_{p} &= \int_0^{\infty} dt\, t\, h(t)\, S_{r}(t).  
\label{eq:p1}
\end{align}
Analogously, we can write 
\begin{align}
1 - p &= \int_0^{\infty} dt\, r(t)\, S_h(t), \nonumber\\[4pt]
(1 - p)\,\tau_{1-p} &= \int_0^{\infty} dt\, t\, r(t)\, S_h(t).
\label{eq:TR}
\end{align}
For exponential $r(t)$, the survival function satisfies
$
S_r(t) = \tau\, r(t),
$
and
$
S_r(t) = -\tau^2\, \dot r(t).
$
Using these relations, the second equation in (\eqref{eq:TR}) yields
\begin{equation}
(1 - p)\,\tau_{1-p} + p\,\tau_p = (1 - p)\,\tau, 
\label{eq:taur-exp}
\end{equation}
which allows us to rewrite Eq.~\eqref{eq:tau_sr} for the MFPT as
\begin{equation}
\tau_{\mathrm{sr}} = \frac{1 - p}{p}\,\tau.
\label{eq:tau_sr_exp}
\end{equation}
Thus, the expression of $\tau_{sr}$ for the SR protocol involves two parameters $\{p,\tau\}$.

For the targer located at $x=L$, the parameter $p$ can be calculated from the first integral in Eq. (\ref{eq:p1}), 
given that $h(t)$ is represented by the inverse Gaussian distribution, $h(t) = \sqrt{L^2/4\pi D t^3} e^{-L^2/4 D t}$,
leading to 
\be
p = e^{-L/\sqrt{D\tau}}.  
\ee

\subsection{SR-A protocol}

As the first extension of the SR protocol, we consider the protocol with "invisible" return trajectories.  
The only contribution of these trajectories is to increase the duration of a single cycle, which in turn 
increases the resulting MFPT.

Because return trajectories are invisible to the target, the probability of a single return trajectory to find a target is $q=0$, 
and the probability of an unsuccessful return trajectory is $1-q=1$.  
Finally, the mean time of an unsuccessful return trajectory is designated by \( \tau_{1-q} \).    

If a target is found after $n$ attempts, the mean time of that sequence of attempts is given by 
\be
\tau_n = (n - 1)(\tau_{1-p} + \tau_{1-q}) + \tau_p, 
\label{eq:Tn2}
\ee
where $\tau_p$ is due to the last successful attempt, and $ (n - 1)(\tau_{1-p} + \tau_{1-q})$ is due to 
$n-1$ unsuccessful attempts which include contributions of an unsuccessdul outgoing and return trajectory.  
The result in Eq. (\ref{eq:Tn2}) together with $p_n$ defined in Eq.~\eqref{eq:pn}, the formula for $\tau_{sr}$ in 
Eq.~\eqref{eq:tau_sr-gen} yields 
\be
\tau_{\mathrm{sr}} = \tau_p + \frac{1 - p}{p} (\tau_{1-p} + \tau_{1-q}).
\label{eq:Tsr-1A}
\ee
And since \( \langle n \rangle = p^{-1} \), this expression can be written as
\begin{equation}
\tau_{\mathrm{sr}} = \langle n \rangle \left[ p \tau_p + (1 - p)(\tau_{1-p}+  \tau_{1-q}) \right],
\end{equation}
indicating that the MFPT is equal to the average number of generations, \( \langle n \rangle \), 
multiplied by a weighted combination of characteristic times — the mean duration of a successful trajectory, \( \tau_p \), 
and the total mean time associated with a failed attempt, \( \tau_{1-p}  +  \tau_{1-q} \).

For the exponential distribution $r(t)$, we can use the relation in Eq.~\eqref{eq:taur-exp}, which modifies 
the formula in Eq. (\ref{eq:Tn2}) to 
\be
\tau_{sr} =   \frac{1-p}{p}  \left(  \tau  +   \tau_{1-q}   \right).  
\label{eq:Tsr-1B}
\ee

\subsubsection{derivation of the MFPT expression}

The only parameter not previously defined is $\tau_{1-q}$, representing the mean time of an unsuccessful trajectory to reach the origin.  
To calculate $\tau_{1-q}$, we need the mean first passage time to $x=0$ from an initial point $x_0$, designated by $t_{\text{on}}(x_0)$ 
and defined in Eq. (\ref{eq:ton2}).  In addition, we need the distribution of initial positions $x_0$, which is different from $n_{\text{off}}$, 
due to the absorbing boundary conditions at the location of a target $x=L$, to account for absence of successful trajectories.  We represent 
such a distribution using the image method involving two Laplace distributions as 
\begin{equation}
n_L(x_0) = A \left[ e^{ - |x_0| / \sqrt{D \tau} } - e^{ - |x_0 - 2L| / \sqrt{D \tau} } \right],
\label{eq:rhoL}
\end{equation}
where \( A \) is a normalization constant such that 
\(
\int_{-\infty}^{L} dx_0\, n_L(x_0) = 1.
\)
The time \(\tau_{1-q} \) can now be represented as 
\be
 \tau_{1-q} = \int_{-\infty}^{L} dx_0\, n_L(x_0)\, t_{\mathrm{on}}(x_0).
\label{eq:T-on}
\ee
Together with Eq.~\eqref{eq:Tsr-1B}, this expression fully determines the MFPT for the SR-A protocol.
In this protocol, the presence of an "invisible" return trajectory extends the MFPT.  The MFPT expression involves  
three parameters $\{p,\tau,\tau_{1-q}\}$.

\subsection{SR-B protocol}

In this section we consider a protocol in which return trajectories are allowed to participate in a target search.  
The probability of finding a target by a single return trajectory in this protocol is $q>0$.  

The probability of finding a target by an outgoing trajectory on a $n$-th attempt is given by 
\be
p_n = \left[(1 - p)(1 - q)\right]^{n - 1} p.
\label{eq:pqn}
\ee
Prior to a $n$-th successful outgoing trajectory, there were $n-1$ unsuccessful outgoing and returning trajectories.  
Setting \( q = 0 \) recovers the SR-A case from Eq.~\eqref{eq:pn}.  

Similarly, the probability of finding a target by a returning trajectory on a $n$-th attempt is 
\begin{equation}
q_n = \left[(1 - p)(1 - q)\right]^{n - 1} (1 - p) q.
\label{eq:qpn}
\end{equation}
Prior to the successful $n$-th return trajectory, there are $n$ unsuccessful outgoing trajectories and $n-1$ unsuccessful return trajectories.  

We can verify that the total probability of locating the target — whether during an outgoing or a return trajectory — is properly normalized:
\[
\sum_{n=1}^\infty \left(p_n + q_n\right) = 1.
\]

We next define the mean time of finding a target for a sequence of $n$ trajectories, considering that a
target is found by an outgoing trajectory, 
\begin{equation}
\tau_n^{\text{out}} = (n - 1)(\tau_{1-p} +  \tau_{1-q}) + \tau_p.  
\label{eq:Tn3-p}
\end{equation}
The expression is similar to that in Eq.~\eqref{eq:Tn2}.  The difference lies in that 
\(  \tau_{1-q} \) is defined differently due to the exclusion of return trajectories that find a target.  

An expression for the mean time of finding a target by a sequence of $n$ trajectories, assuming that the target
is located by a return trajectory, is given by 
\begin{equation}
\tau_n^{\text{ret}} = (n - 1) \tau_{1-q}  +  n \tau_{1-p}  +  \tau_q,
\label{eq:Tn3-q}
\end{equation}
where $\tau_q$ is the mean time of finding a target by a single return trajectory.  

Having defined $p_n$, $q_n$, $\tau_n^{\text{out}}$, and $\tau_n^{\text{ret}}$, 
the MFPT of the entire SR-B process, defined as 
\[
\tau_{\mathrm{sr}} = \sum_{n = 1}^{\infty} \left( p_n \tau_n^{\text{out}} + q_n \tau_n^{\text{ret}} \right), 
\]
yields 
\begin{equation}
\tau_{\mathrm{sr}} =
\frac{
p \tau_p
+ (1 - p) \tau_{1-p}
+ (1 - p) q \tau_q
+ (1 - p)(1 - q)  \tau_{1-q}
}{p + q - pq}.
\label{eq:TsrB-1A}
\end{equation}
For an exponential $r(t)$ we can use Eq.~\eqref{eq:taur-exp} Which reduces the above formula to
\begin{equation}
\tau_{\mathrm{sr}} = 
\frac{ (1 - p)\,\tau \;+\; (1 - p)(1 - q)\,\tau_{1-q} \;+\; q(1 - p)\,\tau_q }{ p + q - p q }.
\label{eq:tau-sr-3}
\end{equation}
The MFPT of the SR-B protocol therefore depends on five parameters,
\(
\{\, p,\, q,\, \tau,\, \tau_{1-q},\, \tau_q \,\}.
\)
Three of these parameters,
\(
\{\, q,\, \tau_{1-q},\, \tau_q \,\},
\)
need to be defined.  $\tau_{1-q}$ of the SR-A protocol is not the same as $\tau_{1-q}$ of the current SR-B protocol.

\subsubsection{derivation of the MFPT expression} 

To define these parameters, we first need to describe the physical scenario underlying the SR-B protocol.  
Because the presence of the target $x=L$ and the off-switch at $x=0$ can be represented as two 
adsorbing walls, the situation corresponds to a random walker enclosed by two adsorbing walls at $x=0$ and $x=L$.  
This gives rise to the splitting probability setup, therefore, the missing parameters can be defined 
within that framework.

For a particle in a harmonic potential, initially at \( x_0 \in (0, L) \), the splitting probability \( \pi_L(x_0) \) --- that is, 
the probability of reaching \( x = L \) before \( x = 0 \) --- is known to satisfy the following backward Fokker–Planck equation 
\cite{Redner2001}:
\[
0 = \mu K x_0\, \pi_L'(x_0) - D\, \pi_L''(x_0),
\]
subject to the boundary conditions \( \pi_L(0) = 0 \) and \( \pi_L(L) = 1 \).
The solution is:
\begin{equation}
\pi_L(x_0) = \frac{ \int_0^{x_0} dx\, e^{\mu K x^2 / (2D)} }{ \int_0^L dx\, e^{\mu K x^2 / (2D)} }.
\label{eq:pi-L}
\end{equation}
To compute \( q \), the probability of locating a target by a single return trajectory, we must consider 
the distribution of initial points $x_0$, which are distributed according to \( n_L(x_0) \), 
already defined in Eq.~\eqref{eq:rhoL}. This yields 
\be
q     =      \int_0^{L} dx_0 \, n_L(x_0) \pi_L(x_0).  
\label{eq:q}
\ee

To determine \( \tau_q \), the mean duration of a single successful return trajectory, we consider 
the mean first passage time for a particle starting at \( x_0 \) to reach \( x = L \) without first being absorbed at \( x = 0 \). 
This quantity, designated by $t_L(x_0)$, satisfies the following backward differential equation:
$$
\pi_L(x_0)     =     \mu K x_0 t_L'(x_0)     -     D t_L''(x_0), 
$$
with boundary conditions \( t_L(0) = t_L(L) = 0 \).
The solution is given by 
\ba
t_L(x_0) &=& \frac{1}{D} \int_0^{x_0} dx\,  e^{\mu K x^2 / 2 D}  \int_x^L dx'\, e^{-\mu K x'^2 / 2D} \pi_L(x')  \nonumber\\ 
&-&   \frac{ \pi_{L}(x_0) }{D} \int_0^{L} dx\,  e^{\mu K x^2 / 2 D}  \int_x^L dx'\, e^{-\mu K x'^2 / 2D} \pi_L(x').  \nonumber
\label{eq:tL}
\ea
The quantity $\tau_q$ is then obtained by considering a distribution of initial points $x_0$, 
\be
q \tau_q =    \int_0^{L} dx_0 \, n_L(x_0)  t_L(x_0).  
\label{eq:Tq}
\ee

Finally, to define \( \tau_{1-q} \), the mean duration of a single unsuccessful return trajectory, which is differently defined from $\tau_{1-q}$ in Eq. (\ref{eq:T-on}), 
we consider the mean first passage time for a particle starting at \( x_0 \) to reach \( x = 0 \) without first being absorbed at \( x = L \). 
This quantity, designated by $t_0(x_0)$, satisfies the following differential equation
\begin{equation}
\pi_0(x_0) = \mu K x_0\, t_0'(x_0) - D\, t_0''(x_0),
\end{equation}
where \( \pi_0(x_0) = 1 - \pi_L(x_0) \) is the splitting probability to reach \( x = 0 \) before \( x = L \).
The solution is 
\begin{align}
t_0(x_0) &= \frac{1}{D} \int_{x_0}^{L} dx\, e^{\mu K x^2 / (2D)} \int_0^x dx'\, e^{-\mu K x'^2 / (2D)} \pi_0(x') \nonumber\\
&\quad - \frac{ \pi_0(x_0) }{D} \int_0^L dx\, e^{\mu K x^2 / (2D)} \int_0^x dx'\, e^{-\mu K x'^2 / (2D)} \pi_0(x').
\label{eq:t-0}
\end{align}
The quantity \(\tau_{1-q} \) is then obtained from the relation 
\be
(1-q) \tau_{1-q}   =  
 \int_0^{L} dx_0 \, n_L(x_0)  t_0(x_0)    +    \int_{-\infty}^{0} dx_0\, n_L(x_0)  t_{\text{on}}(x_0), 
\label{eq:Tbar}
\ee
where the second integral accounts for the initial positions $x_0<0$ that are prevented from reaching the target at $x=L$ due to the adsorbing wall 
at $x=0$.  We recall that $t_{\text{on}}$ is defined in Eq. (\ref{eq:ton2}).

\subsubsection{results} 

\begin{figure}[hhhh] 
 \begin{tabular}{rrrr}
\includegraphics[height=0.23\textwidth,width=0.25\textwidth]{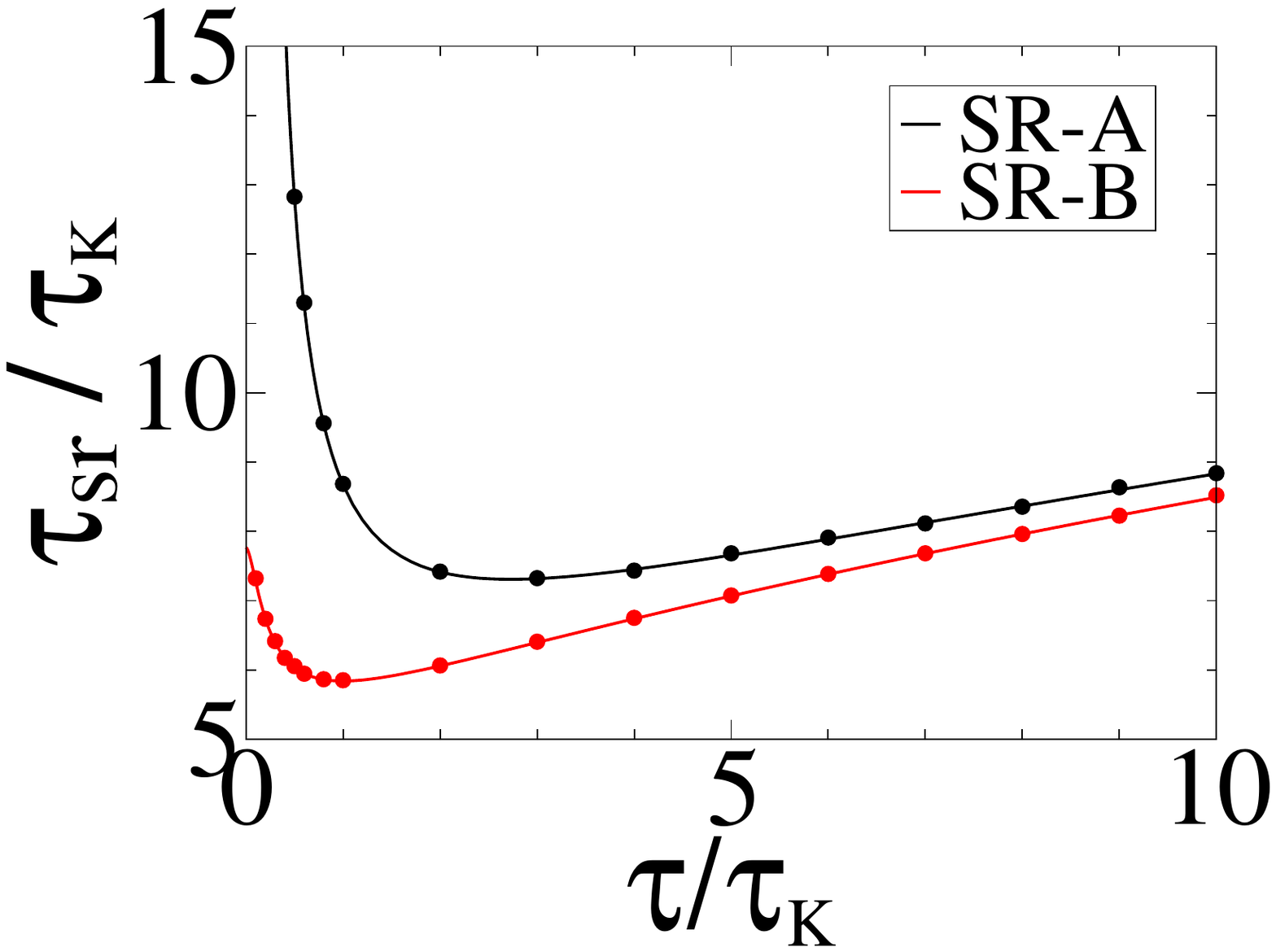}  
 \end{tabular}
\caption{MFPT as a function of $\tau$ for the SR-A and SR-B protocols, corresponding to Eq. (\ref{eq:Tsr-1B}) and Eq. (\ref{eq:tau-sr-3}), 
compared with the simulation data points.  The results are for $b=1/\sqrt{0.3}$, where $b=L/\sqrt{D\tau_K}$.  }
\label{fig:tau-sr-sim} 
\end{figure}

Having defined all the parameters, we are ready to calculate the MFPT.  
Fig.~(\ref{fig:tau-sr-sim}) compares the MFPTs for the SR-A and SR-B protocols, given by Eqs.~\eqref{eq:Tsr-1B} and \eqref{eq:tau-sr-3}, 
with the data points from numerical simulations.  The close agreement across the explored parameter range confirms the correctness of the analytical formulas.
\graphicspath{{figures/}}

Since both SR-A and SR-B protocols include return trajectories, their MFPTs are expected to be greater than that of the SR protocol. 
And since in SR-B the return trajectories contribute to the search, the overhead time associated with the return motion is reduced. 
Consequently, we anticipate the following relation:
\(
\tau_{\mathrm{sr}}^{\mathrm{SR}} \;\leq\; \tau_{\mathrm{sr}}^{\mathrm{SR\text{-}B}} \;\leq\; \tau_{\mathrm{sr}}^{\mathrm{SR\text{-}A}}.
\)
\graphicspath{{figures/}}
\begin{figure}[hhhh] 
 \begin{tabular}{rrrr}
\includegraphics[height=0.21\textwidth,width=0.24\textwidth]{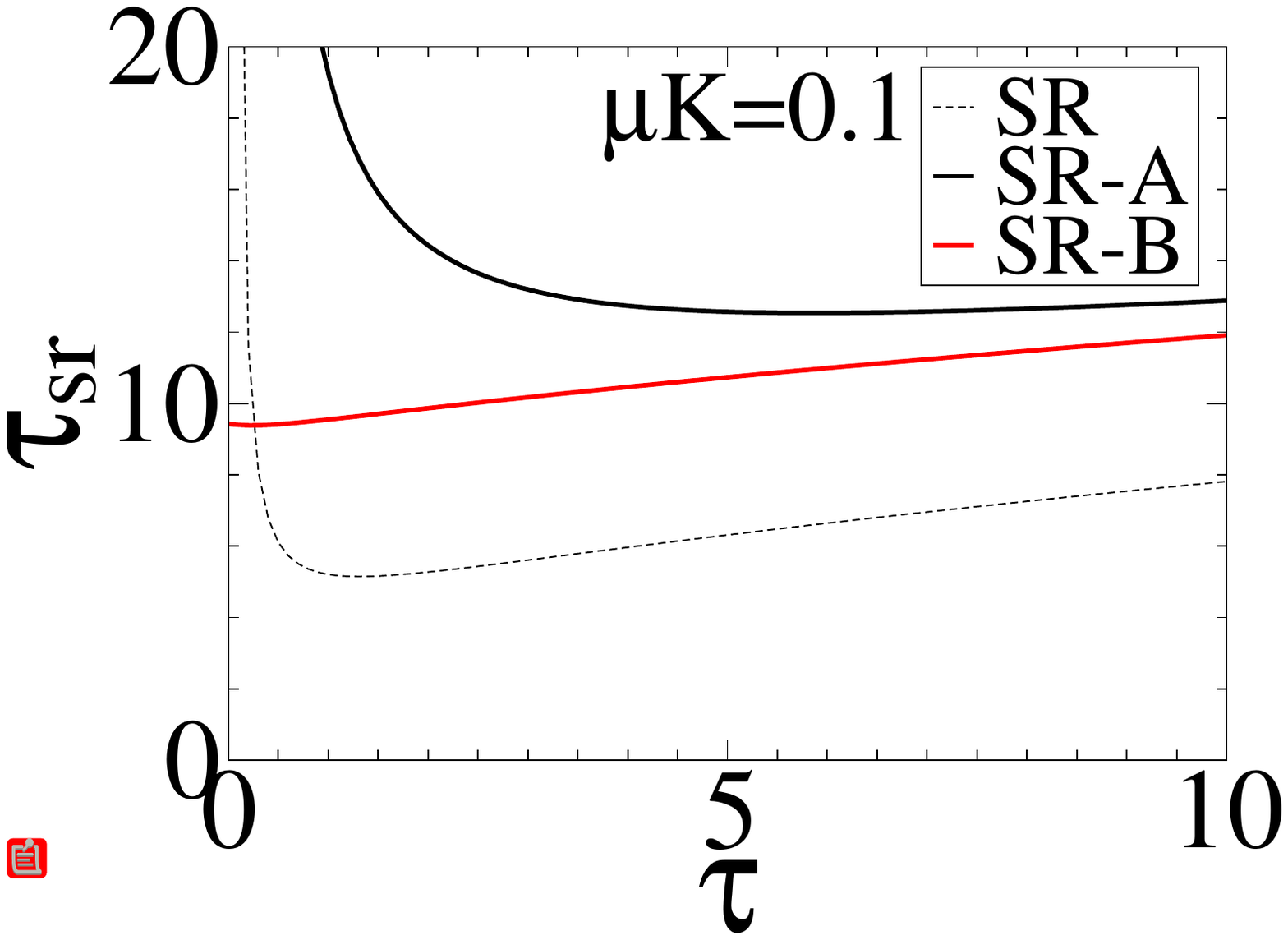}  \\
\includegraphics[height=0.21\textwidth,width=0.24\textwidth]{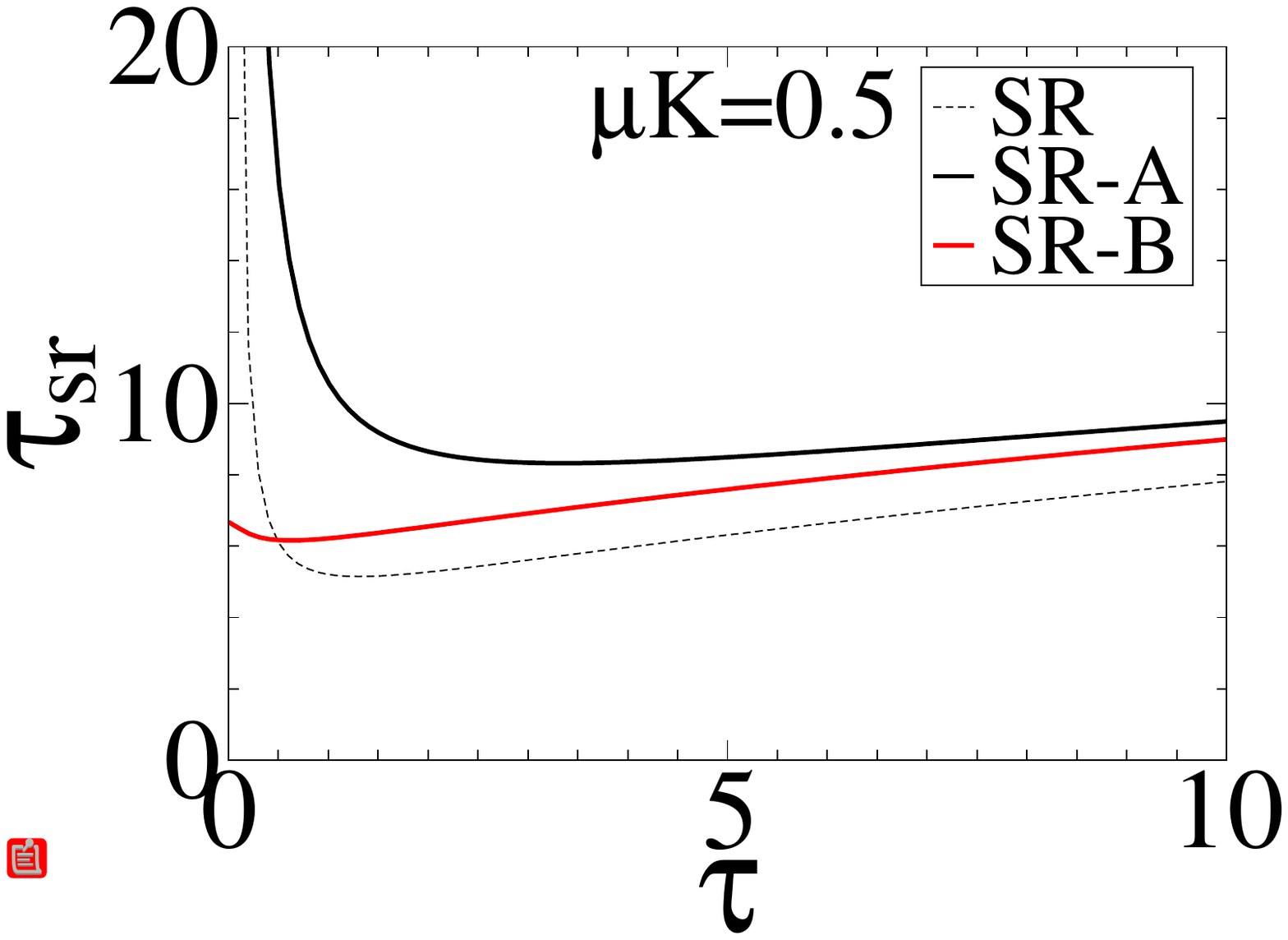}  \\
\includegraphics[height=0.21\textwidth,width=0.24\textwidth]{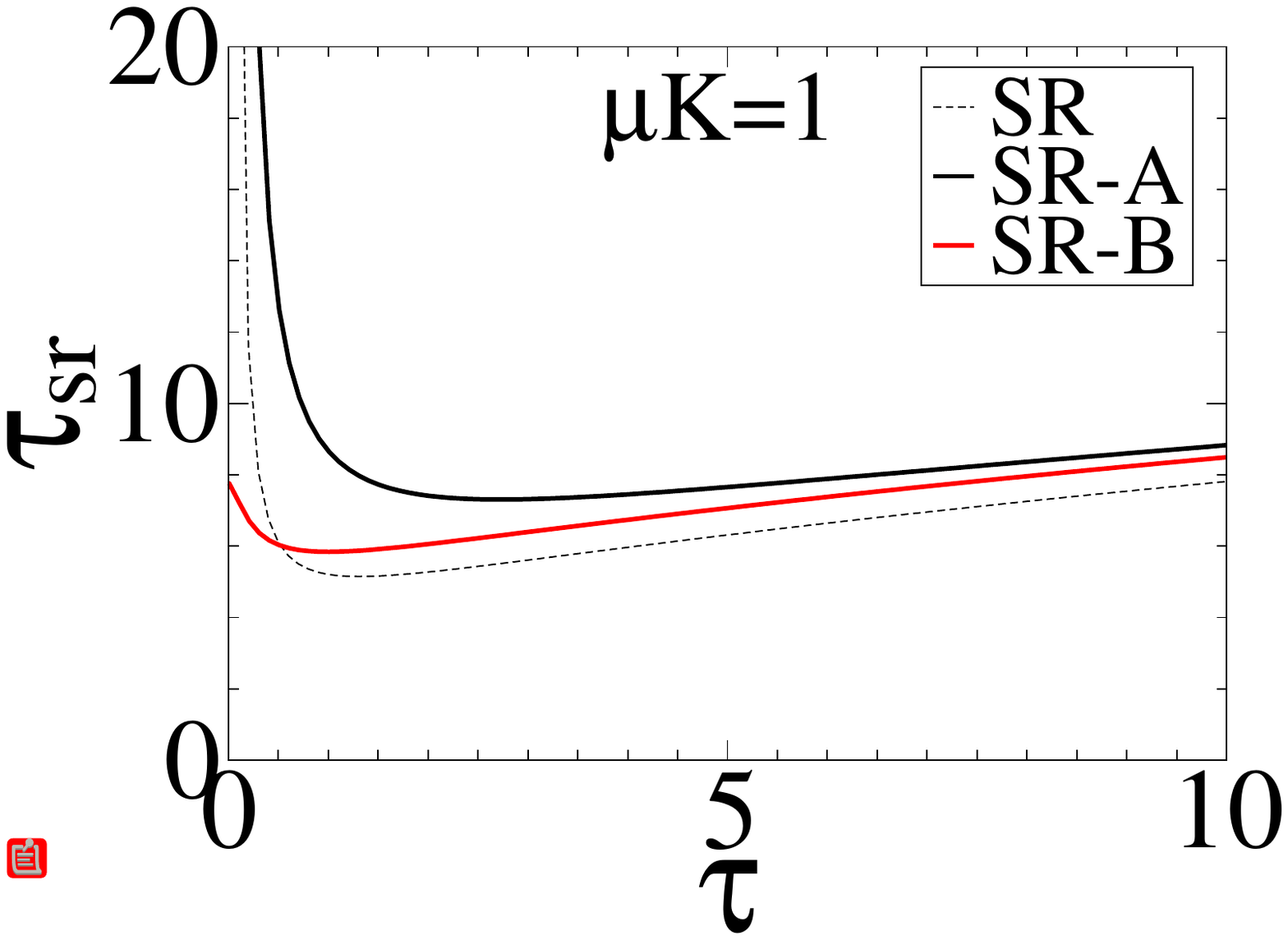}  
 \end{tabular}
\caption{MFPT as a function of $\tau$ for three types of the SR protocol and for three values of the stiffness parameter, $K$.  
The results in this figure are for $L=1$ and $D=0.3$.    }
\label{fig:tau-sr} 
\end{figure}

In Fig.~\ref{fig:tau-sr}, we plot the MFPT as a function of $\tau$ for the three protocols analyzed above.  
Note that the relation 
\(
\tau_{\mathrm{sr}}^{\mathrm{SR}} \;\leq\; \tau_{\mathrm{sr}}^{\mathrm{SR\text{-}B}} \;\leq\; \tau_{\mathrm{sr}}^{\mathrm{SR\text{-}A}}
\)
is satisfied most of the time, except in the region of small $\tau$, where SR-B becomes the fastest protocol.  
It is also the only protocol whose MFPT does not diverge in the limit $\tau \to 0$.  
This suggests that reverse search under certain conditions becomes more efficient than the forward search.

To understand this situation, we need to clarify why the MFPT diverges for the SR and SR-A protocols as \(\tau \to 0\).  
In these protocols, the outgoing trajectories become too short-lived to reach the target.  
As a result, the number of generations required for a successful hit diverges.  

In the \(\tau \to 0\) limit, outgoing trajectories are effectively suppressed.  
This implies that only return trajectories can contribute to locating the target.  
However, for SR and SR-A, return trajectories are either absent or invisible to the target.  
Only the SR-B protocol can use these return trajectories productively in the search.

Given the vanishing duration of the "off" state in the limit \(\tau \to 0\),  
the harmonic trap is effectively permanent, and the initial position of each return trajectory approaches the origin.  
The problem reduces to computing the MFPT for a particle starting at \(x = 0\) in a harmonic potential,  
searching for a target located at \(x = L\).
The MFPT for this problem is well known~\cite{Redner2001,Chelminiak2024} and given by:
\begin{equation}
\tau_{sr}^{0}
= \frac{1}{D} \int_{0}^{L} dx\, 
e^{{\mu K} x^{2}/ {2D}}
 \int_{-\infty}^{x} dx' \,
e^{-{\mu K} x'^{2}/ {2D}},
\end{equation}
which evaluates to
\begin{equation}
\tau_{\mathrm{sr}}^0
=  
\frac{\tau_K}{2} \left[ \pi \operatorname{erfi} \sqrt{\tfrac{b^2}{2}} 
\;+\;
b^{2}\; {}_2F_2\!\left(1,1;\tfrac{3}{2},2;\tfrac{b^{2}}{2}\right)\right],
\label{eq:srB-tau0-limit}
\end{equation}
where \( b = L / \sqrt{D \tau_K} \), and \({}_2F_2\) denote generalized hypergeometric functions.

This expression, though finite, is not bounded in \(b\); it grows with distance to the target.  
But critically, it demonstrates that the MFPT for SR-B remains finite as \(\tau \to 0\), unlike the diverging behavior of SR and SR-A.

We conclude that the enhanced efficiency of SR-B in the \(\tau \to 0\) regime stems not from the intrinsic advantage of reverse search,  
but from the fact that it remains operative when outgoing trajectories are suppressed.  It reduces to the first passage problem of a 
particle in a permanent trap.  This behavior has been observed in \cite{PRE-Arnab-2024,PRE-Arnab-2025}, which primarily focused 
on a linear potential.

%
%


\subsection{SR-C:  protocol without the outgoing trajectories}

In this section, we introduce a protocol that entirely eliminates outgoing trajectories: the search is performed 
solely via return paths. In this scenario, a particle begins its motion from a position sampled from the distribution 
\( n_{\text{off}}(x_0) \), given in Eq. (\ref{eq:Laplace}), 
and is then driven by a harmonic force toward the origin. If the particle reaches 
the origin without locating the target, it is instantaneously relocated to a new starting position, again drawn from 
\( n_{\text{off}} \). This instantaneous (and unphysical) reset resembles the standard SR protocol, but with a critical distinction.

Unlike traditional SR, the present model incorporates an additional control parameter—the strength of the restoring force—
that allows us to modulate the efficiency of the search. Furthermore, the driven return motion bears similarities to active 
motion. However, in contrast to active motion, where directionality is stochastic, here the motion is 
persistently oriented toward the origin.

\subsubsection{time-dependent distribution}

Since the idea of stochastic resetting suggested by the SR-C protocol is novel, in this section we analyze its 
time-dependent distribution, \(n_{\text{on}}(x,t)\), which satisfies the partial differential equation:  
\begin{equation}
\dot n_{\text{on}} = \left[ \mu K x\, n_{\text{on}} \right]' + D\,  n''_{\text{on}} + j(t)\, n_{\text{off}}(x),
\end{equation}
with boundary condition \( n_{\text{on}}(0,t) = 0 \), and initial condition \( n_{\text{on}}(x,0) = n_{\text{off}}(x) \).  
The flux at the absorbing boundary is defined by
\begin{equation}
j(t) = D\, \partial_x n_{\text{on}}(x,t)\big|_{x=0}.
\end{equation}
We consider only the distribution in the half-line \( x \geq 0 \), which is sufficient due to the even symmetry:
\(
n_{\text{on}}(x,t) = n_{\text{on}}(-x,t).
\)

The solution in the positive half-space admits the renewal form:
\begin{align}
n_{\text{on}}(x,t) &= \int_0^{\infty} dx'\, G_a(x,x',t)\, n_{\text{off}}(x') \nonumber\\
&\quad + \int_0^t dt'\, j(t') \int_0^{\infty} dx'\, G_a(x,x',t - t')\, n_{\text{off}}(x'). 
\label{eq:n_on-general}
\end{align}
The first term describes evolution of the initial state, while the second accounts for successive reinjections at the origin.

The flux function \( j(t) \) satisfies another renewal equation \cite{PRE-PalReuveni-2019,PRL-PalReuveni-2018,JCP-RenewalFlux-2023}, 
\begin{equation}
j(t) = r_{\text{on}}(t) + \int_0^t dt'\, j(t')\, r_{\text{on}}(t - t'),
\end{equation}
which captures the recursive structure of returns.  At short times, \( j(t) \approx r_{\text{on}}(t) \), while at long times 
the flux approaches its steady-state value:
\[
j(t) \to \tau_{\text{on}}^{-1}.
\]
It can be verified that the distribution is conserved at all times, 
\begin{equation}
\int_0^{\infty} dx\, n_{\text{on}}(x,t) = \tfrac{1}{2}.
\end{equation}
As \( t \to \infty \), the distribution \( n_{\text{on}}(x,t) \) approaches the stationary form given by Eq.~(\ref{eq:n-on-integral}).

\subsubsection{MFPT}

Based on previous expressions of $\tau_{sr}$ derived for other protocols, the MFPT of the SR-C protocol is found to be 
defined analogously to Eq.~(\ref{eq:tau_sr}):
\begin{equation}
\tau_{sr} = \frac{ q\, \tau_q + (1-q)\, \tau_{1-p} }{q},
\label{eq:tau-sr-d}
\end{equation}
where \( q \) is the probability that a single (return) trajectory finds the target, \( \tau_q \) is the mean duration 
of a single successful trajectory, and \( \tau_{1-p} \) is the mean duration of a single unsuccessful trajectory (one that reaches 
\( x = 0 \) without finding the target). Thus, the MFPT depends on three parameters, 
\( \{ q, \tau_q, \tau_{1-p} \} \), which we will define below.

\subsubsection{derivation of the MFPT expression}

We define the probability for a single trajectory to find the target as  
\be
q = \int_{L}^{\infty} dx_0\, n_{\text{off}}(x_0)  +    \int_{0}^{L} dx_0\, n_{\text{off}}(x_0)  \pi_L(x_0). 
\label{eq:qc-0}
\ee
The first term of the expression captures contributions of all the trajectories with the initial position $x_0\geq L$.  
All these trajectories find the target en route to $x=0$, assuming that $K>0$.   
The second term represents the contributions of the trajectories with the initial position in the interval $x_0\in (0,L)$.  
These trajectories find the target with 
the probability $\int_{0}^{L} dx_0\, n_{\text{off}}(x_0)  \pi_L(x_0)$, where $\pi_L(x_0)$ is the splitting probability defined in Eq. (\ref{eq:pi-L}).  
After evaluating the integral, Eq. (\ref{eq:qc-0}) becomes 
\be
q = 
\frac{ e^{-\alpha/2} }{ 2 } \, 
\frac{
\operatorname{erfi}\! \sqrt{\tfrac{\alpha}{2}} 
- \operatorname{erfi}\!\left( \sqrt{\tfrac{\alpha}{2}} - \sqrt{\tfrac{b^{2}}{2}} \right)  }
{\operatorname{erfi}\! \sqrt{\tfrac{b^{2}}{2}} },
\label{eq:qc-1}
\ee
where $b=L/\sqrt{D\tau_K}$ and $\alpha = \tau_K/\tau$.

The mean time of a single unsuccessful trajectory to reach $x=0$, designated by $\tau_{1-p}$, is obtained 
from the following relation 
\be
(1-q) \tau_{1-p}   =  
 \int_0^{L} dx_0 \, n_{\text{off}}(x_0)  t_0(x_0)    +    \int_{-\infty}^{0} dx_0\, n_{\text{off}}(x_0)  t_{\text{on}}(x_0), 
\label{eq:Tbbar}
\ee
where $t_0(x_0)$ and $t_{\text{on}}(x_0)$ have been defined and discussed previously.  
The second term represents trajectories that are prevented from 
reaching the target due to the location of the off switch at $x=0$.  

The mean time of a single successful trajectory is obtained from 
\be
q\tau_q = \int_{L}^{\infty} dx_0\, n_{\text{off}}(x_0) t_{q}(x_0)  +    \int_{0}^{L} dx_0\, n_{\text{off}}(x_0)  t_L(x_0). 
\ee
The expression distinguishes between different contributions.  The first term represents trajectories with an initial 
point $x_0\geq L$.  All these trajectories will eventually find a target.   The quantity 
$t_q(x_0)$ is the mean time to reach $x=L$ for trajectories that start at $x_0$.   It satisfies the following differential equation \cite{Redner2001}
$$
1     =     \mu K x_0 t_q'(x_0)     -     D t_q''(x_0), 
$$
where the solultion for the boundary conditions $t_q(L) = 0$ is 
\be
t_q(x_0) = \frac{1}{D} \int_{L}^{x_0} dx\,  e^{\mu K x^2 / 2 D}  \int_x^{\infty} dx'\, e^{-\mu K x'^2 / 2D}.  
\ee
The second term represents the trajectories with an initial point located in the interval $x_0\in (0,L)$.  
The quantity $t_{L}(x_0)$ was already defined in Eq. (\ref{eq:tL}).

After evaluating all the integrals involved in Eq.~(\ref{eq:tau-sr-d}), we obtain the following compact expression in terms 
of previously defined quantities:
\be
\tau_{sr} =   \frac{\tau_{\text{on}} }{q}  - t_{\text{on}}(L),
\label{eq:tsr-src}
\ee
where $t_{\text{on}}(x_0)$ is defined in Eq.~(\ref{eq:ton2}), $\tau_{\text{on}}$ in Eq.~(\ref{eq:tau-on-0-2}), and $q$ is
given by the formula in Eq. (\ref{eq:qc-1}).


The simplicity and elegance of Eq.~(\ref{eq:tsr-src}) invite physical interpretation. 
Both time scales describe the mean time a particle needs to reach the origin. The first term, $\tau_{\text{on}}$, is the average 
over all possible initial positions $x_0$, while the second term, $t_{\text{on}}(L)$, corresponds to the mean time from the  
position of the target, $x_0 = L$. 
It would be essentially impossible to deduce this relation from physical intuition alone. Its unusual structure strongly suggests 
that it arises from a sequence of mathematical cancellations within the derivation.

\subsubsection{results} 

\begin{figure}[t] 
  \centering
  \includegraphics[height=0.23\textwidth,width=0.25\textwidth]{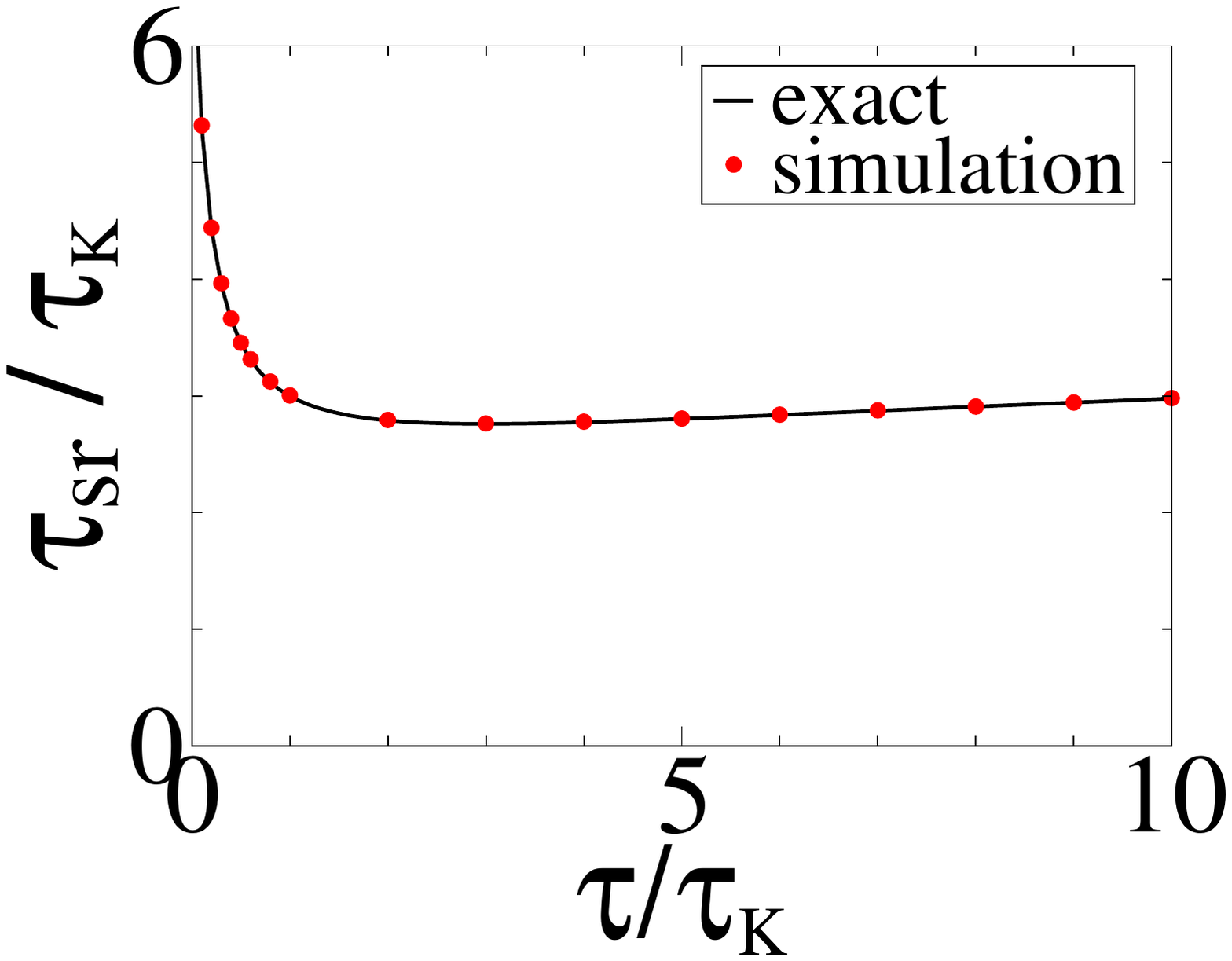}
  \caption{MFPT as a function of \(\tau\) for the SR-C protocol, comparing the analytical expression 
  from Eq.~(\ref{eq:tau-sr-d}) with data from numerical simulations. Parameters: \(b = L/\sqrt{D\tau_K} = 1/\sqrt{0.3}\).}
  \label{fig:th-d} 
\end{figure}

To confirm the analytical expressions, in Fig.~(\ref{fig:th-d}) we plot the MFPT 
obtained from Eq.~(\ref{eq:tsr-src}) and compare it with numerical simulations. The close agreement between 
the two confirms the validity of the theoretical framework.
\graphicspath{{figures/}}

\begin{figure}[t]
  \centering
  \begin{tabular}{c}
    \includegraphics[height=0.21\textwidth,width=0.24\textwidth]{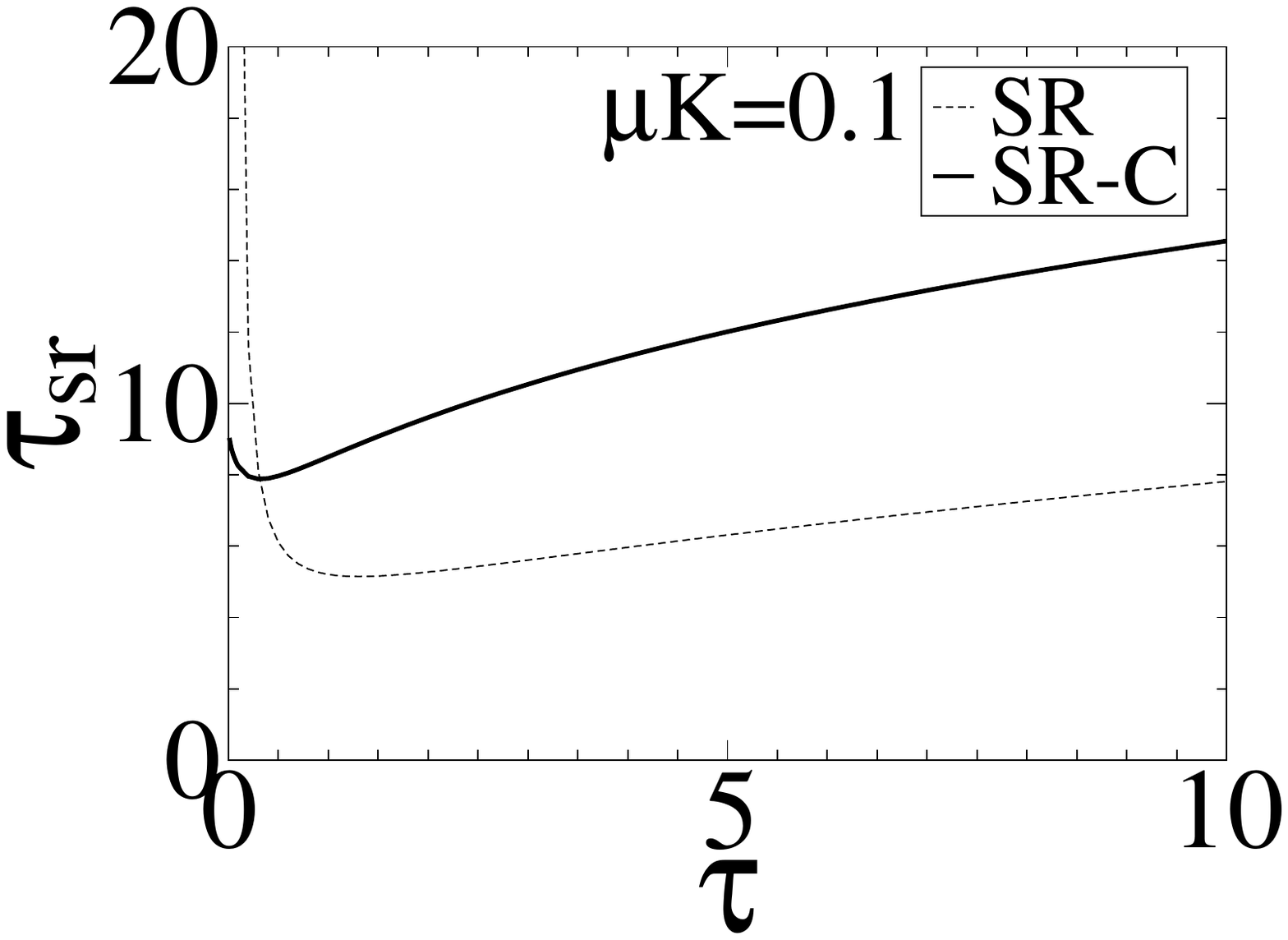} \\
    \includegraphics[height=0.21\textwidth,width=0.24\textwidth]{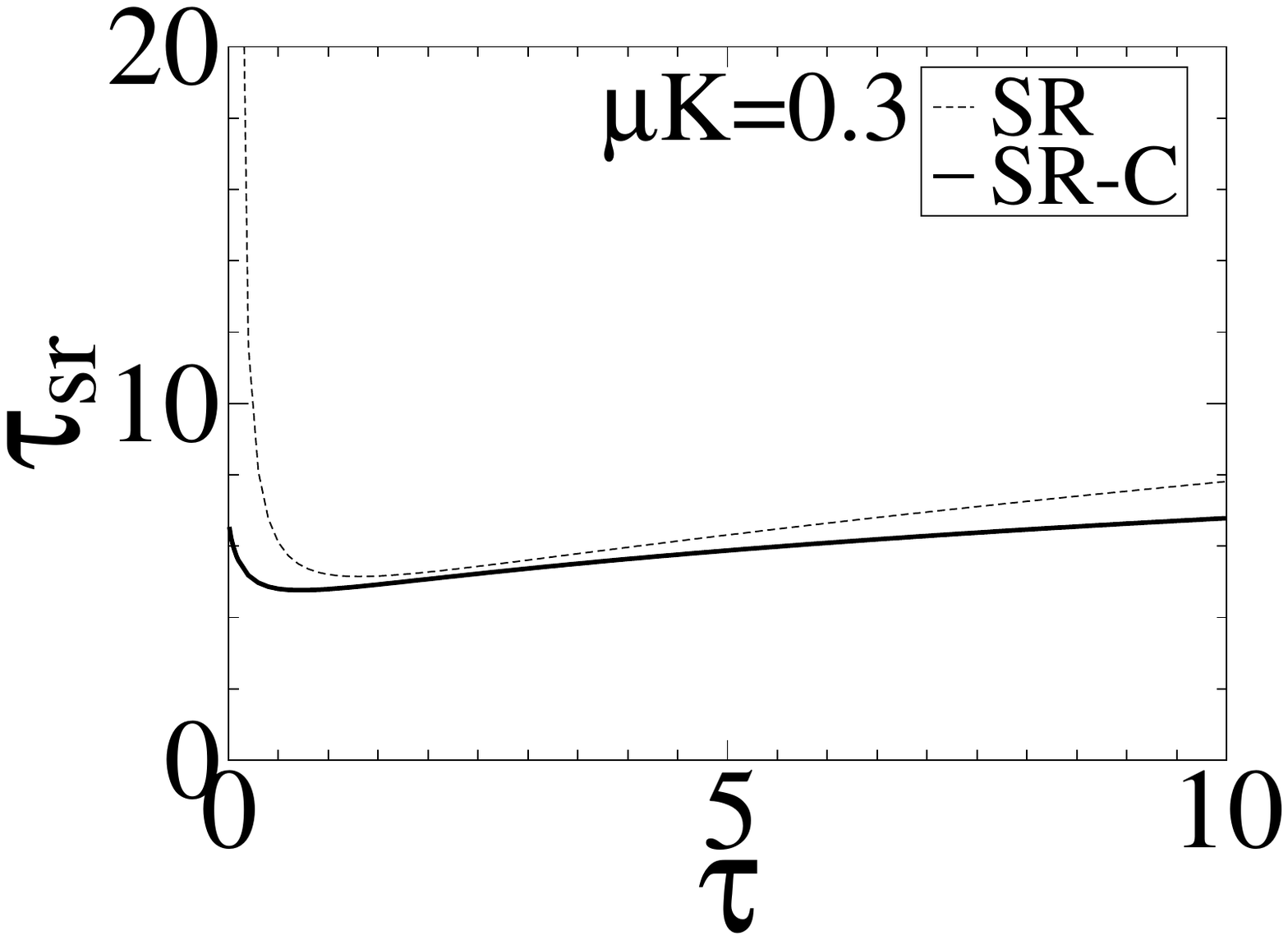} \\
    \includegraphics[height=0.21\textwidth,width=0.24\textwidth]{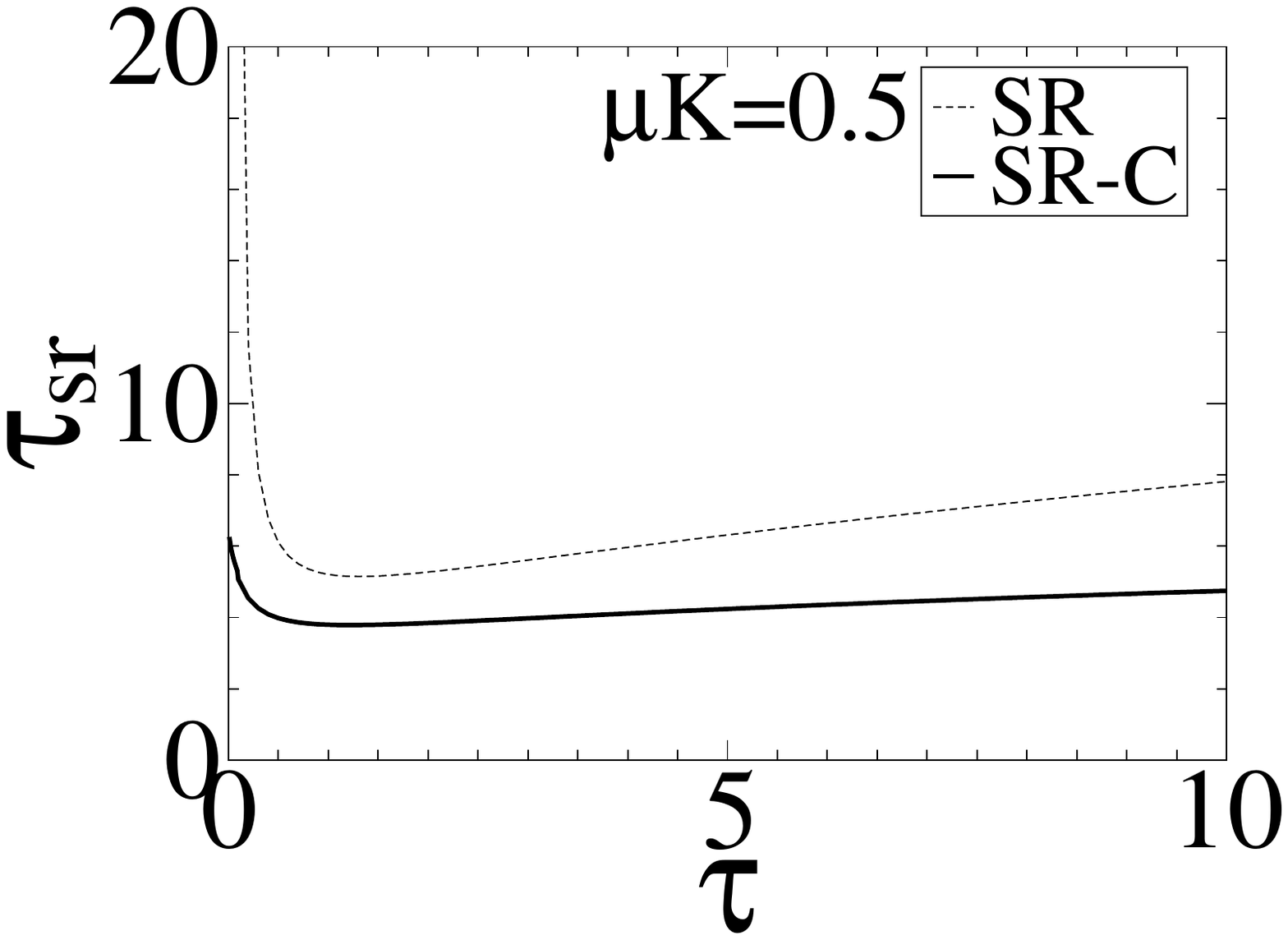}
  \end{tabular}
  \caption{MFPT as a function of \(\tau\) for the SR-C protocol (solid lines) compared with the SR 
  protocol (dashed lines), for three values of stiffness \(K\). Parameters: \(L = 1\), \(D = 0.3\).}
  \label{fig:tau-sr-d}
\end{figure}

In Fig.~(\ref{fig:tau-sr-d}), we show \(\tau_{sr}\) as a function of the waiting time \(\tau\) for several values 
of the trap stiffness \(K\), comparing the SR-C and SR protocols. For small \(K\), the SR protocol yields 
a lower MFPT, i.e., \(\tau_{\mathrm{sr}}^{\mathrm{SR}} < \tau_{\mathrm{sr}}^{\mathrm{SR\text{-}C}}\), 
except in the regime of small \(\tau\), where \(\tau_{\mathrm{sr}}^{\mathrm{SR}}\) diverges while 
\(\tau_{\mathrm{sr}}^{\mathrm{SR\text{-}C}}\) approaches a finite limit, consistent with Eq.~(\ref{eq:srB-tau0-limit}). 
As \(K\) increases, however, the SR-C protocol becomes more efficient and eventually outperforms SR-B across the entire range of \(\tau\).
\graphicspath{{figures/}}

\begin{figure}[t] 
  \centering
  \includegraphics[height=0.21\textwidth,width=0.25\textwidth]{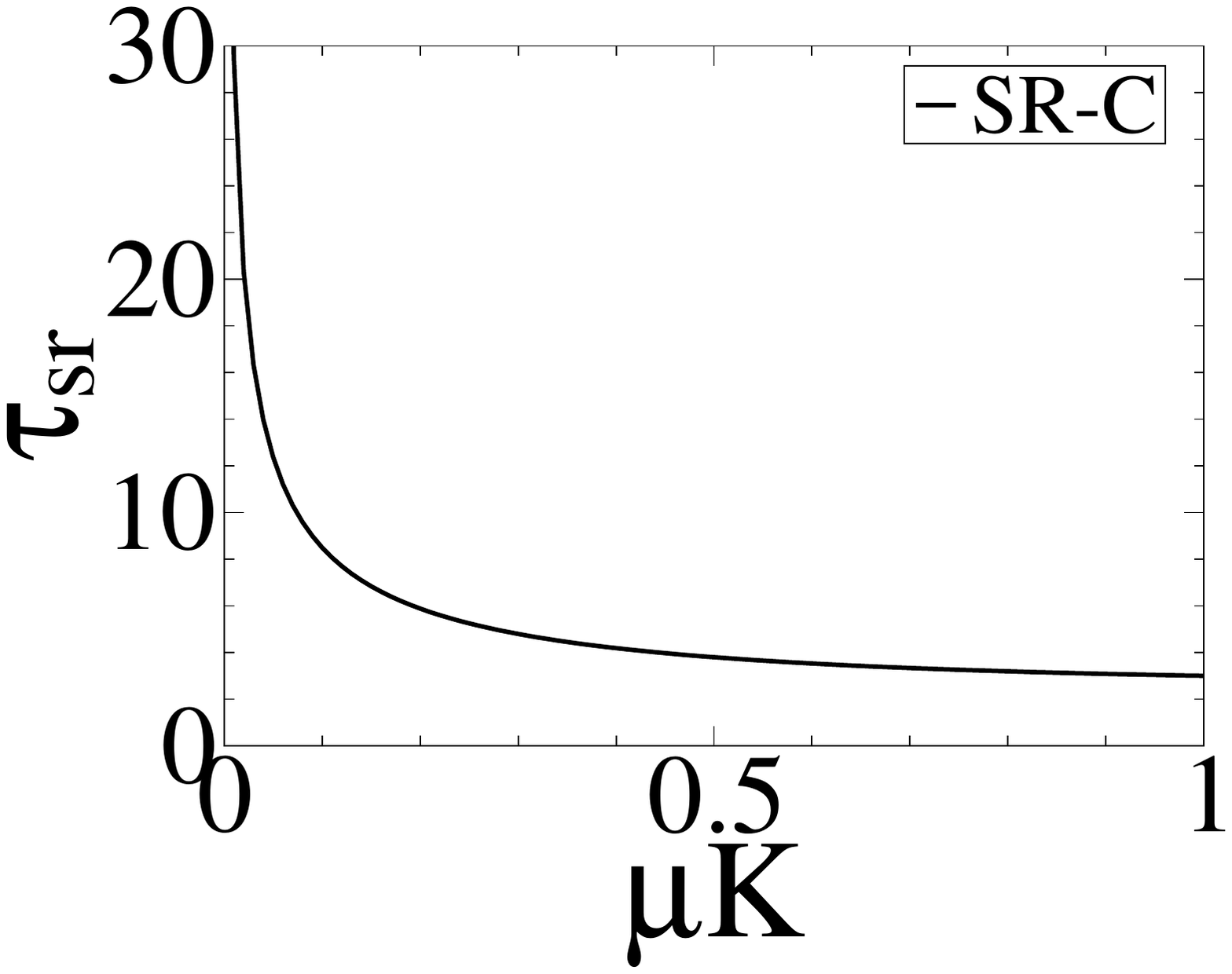}
  \caption{MFPT \(\tau_{sr}\) as a function of trap stiffness \(K\) for the SR-C protocol. Parameters: \(\tau = 1\), \(L = 1\), \(D = 0.3\).}
  \label{fig:tsr-src-K} 
\end{figure}

The dependence of \(\tau_{sr}\) on trap stiffness \(K\) is further illustrated in Fig.~(\ref{fig:tsr-src-K}), where \(\tau_{sr}\) 
is plotted for a fixed \(\tau = 1\). Since both \(\tau\) and \(D\) are held constant, the initial distribution of return positions \(x_0\) 
is independent of \(K\), following \(n_{\text{off}}(x_0) \propto \exp(-|x_0|/\sqrt{D\tau})\). Thus, any variation in MFPT arises 
purely from the dynamics of the return trajectories. As \(K\) increases, particles are drawn more strongly toward the center 
by the trap, resulting in shorter return times and a more efficient search process.
\graphicspath{{figures/}}

\begin{figure}[t] 
  \centering
  \includegraphics[height=0.21\textwidth,width=0.25\textwidth]{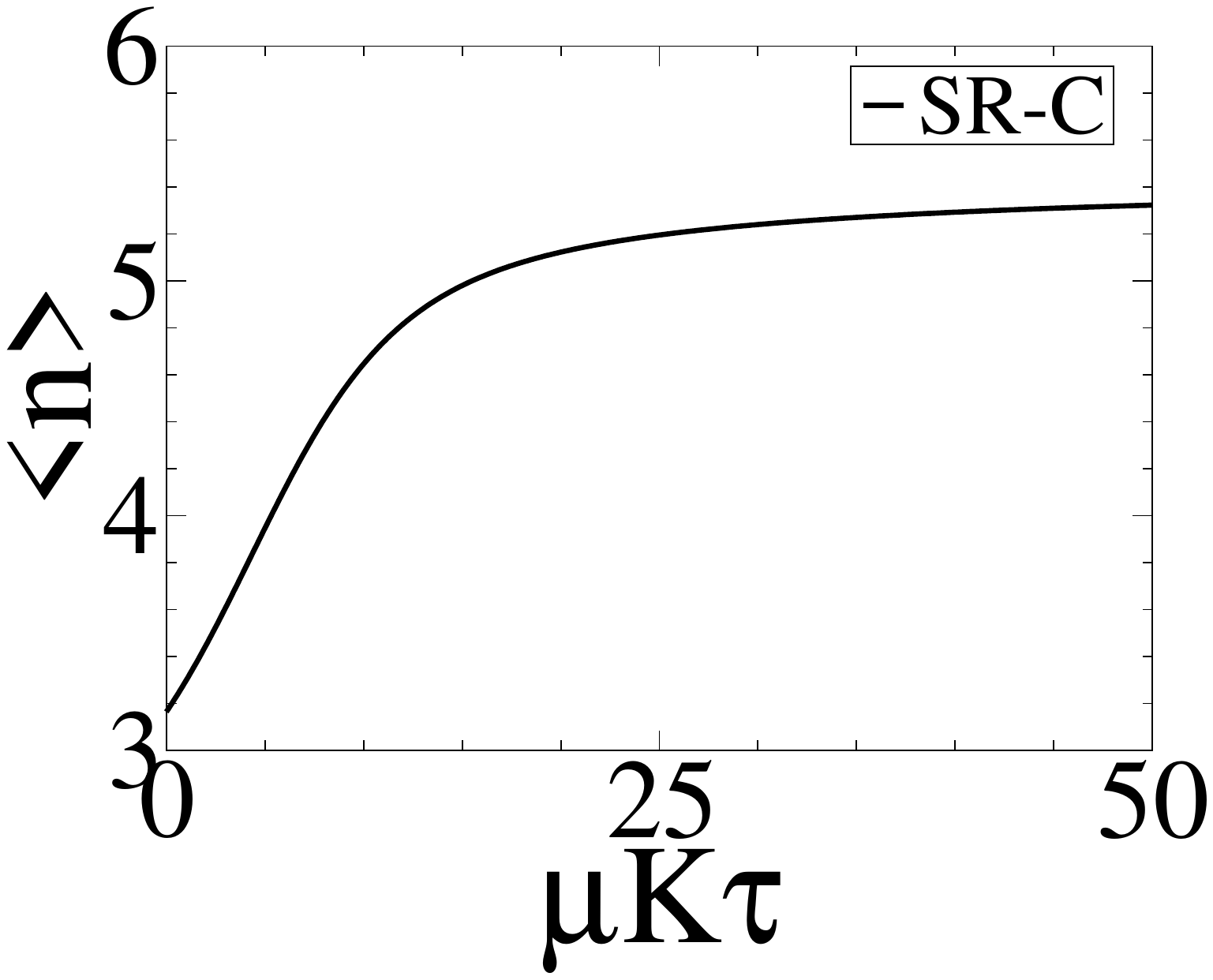}
  \caption{Mean number of generations \(\langle n \rangle = q_c^{-1}\) required to find the target, 
  as a function of trap stiffness \(K\). Parameters: \(\tau = 1\), \(L = 1\), \(D = 0.3\).}
  \label{fig:n-ave-src} 
\end{figure}

To better understand the mechanism underlying this reduction in \(\tau_{sr}\), Fig.~(\ref{fig:n-ave-src}) shows the mean 
number of generations required to locate the target, \(\langle n \rangle = q^{-1}\), as a function of \(K\). 
At first glance, it may seem paradoxical that \(\langle n \rangle\) increases with stiffness—suggesting a less efficient search—
seemingly contradicting the behavior observed in Fig.~(\ref{fig:tsr-src-K}).
\graphicspath{{figures/}}

This apparent contradiction is resolved by analyzing the structure of the initial distribution \(n_{\text{off}}(x)\), 
which divides the population of return trajectories into two groups: a fraction \(\int_L^\infty dx\, n_{\text{off}}(x)\) 
originates on the "far side" of the target (\(x_0 > L\)), and the remainder on the "near side" (\(x_0 < L\)). As \(K\) increases, 
the steeper potential makes it harder for particles from \(x_0 < L\) to reach the target, thus reducing their success probability. 
In contrast, all particles from \(x_0 > L\) are deterministically pulled toward the target and will eventually cross it. 

\begin{figure}[t] 
  \centering
  \includegraphics[height=0.21\textwidth,width=0.25\textwidth]{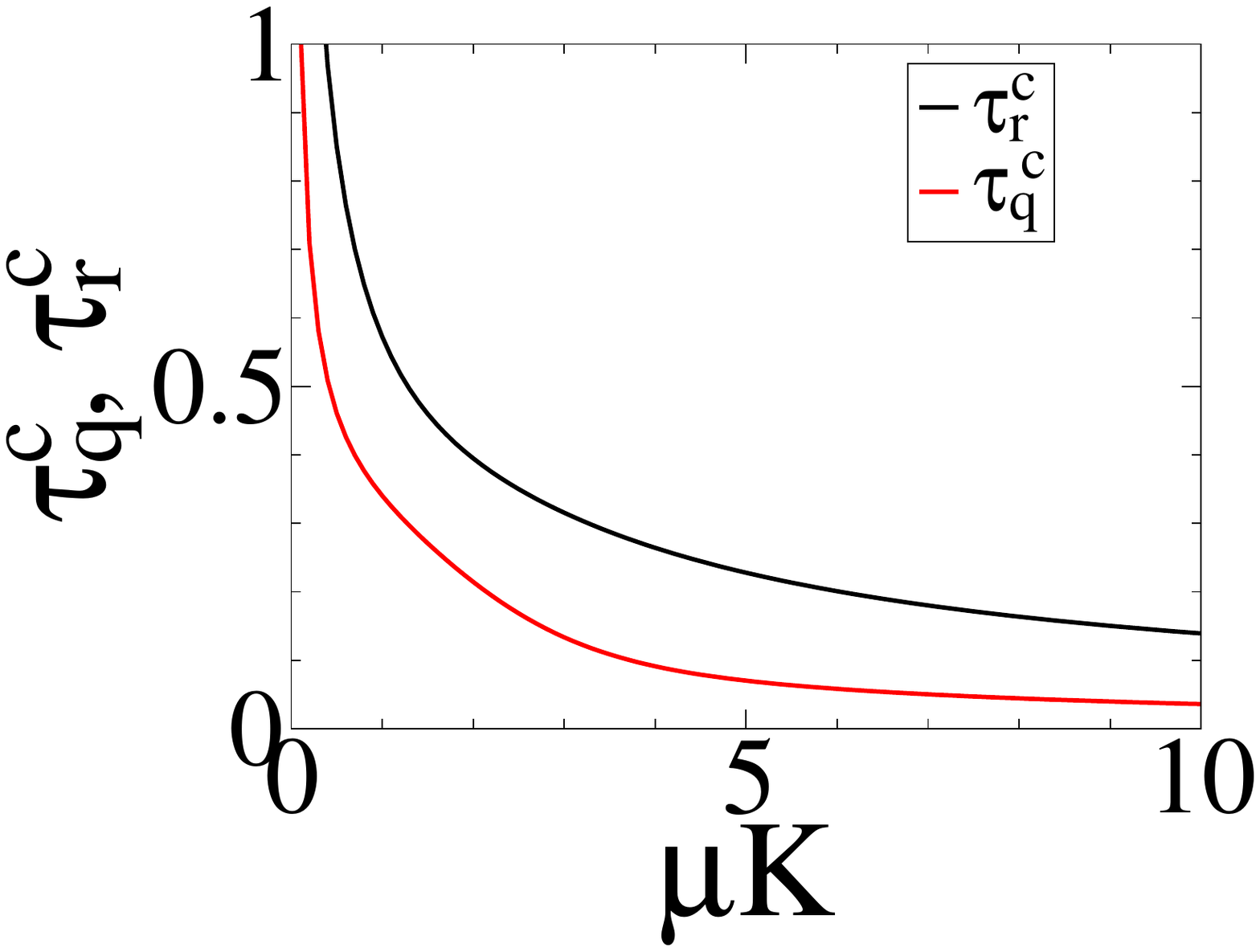}
  \caption{The success probability \(q_c\) as a function of trap stiffness \(K\) for the SR-C protocol.  
  Parameters: \(L/\sqrt{D\tau} = 1\).}
  \label{fig:q-src} 
\end{figure}

Fig.~(\ref{fig:q-src}) confirms that although \(\langle n\rangle = q^{-1}\) increases with trap stiffness \(K\), 
the two characteristic time scales of the SR-C protocol, \(\tau_q\) and \(\tau_{1-p}\), both decrease. 
This explains the overall reduction of the MFPT with increasing \(K\) as due to the greater search efficiency 
of particles originating on the "far side" of the target (\(x_0 > L\)), which are driven deterministically toward the target by the stronger restoring force.
\graphicspath{{figures/}}

\begin{figure}[t] 
  \centering
  \includegraphics[height=0.19\textwidth,width=0.23\textwidth]{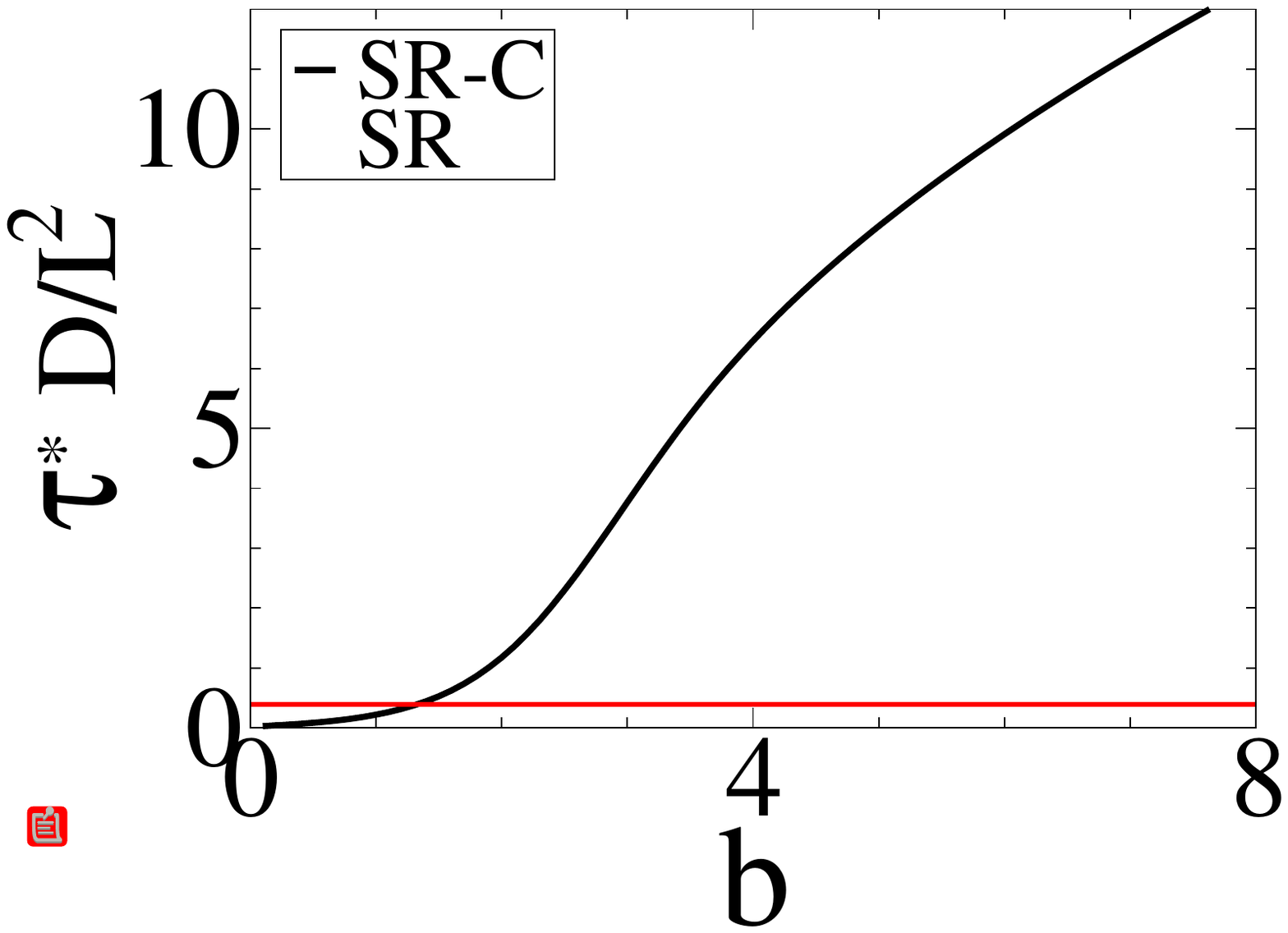}
  \includegraphics[height=0.19\textwidth,width=0.23\textwidth]{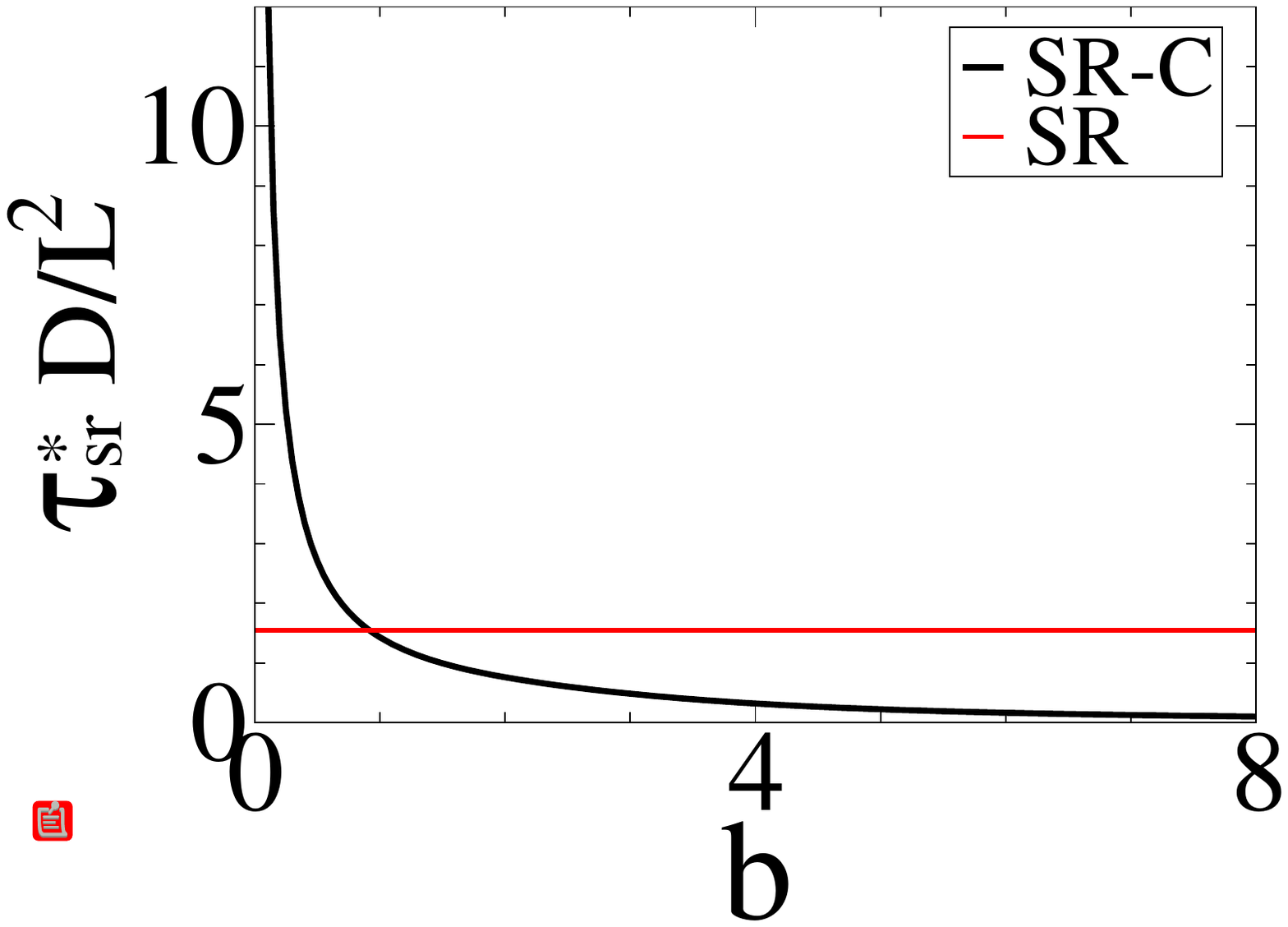}
  \caption{
  The left panel shows the optimal reset time \( \tau^* \), corresponding to the minimum of \( \tau_{\text{sr}} \), 
  as a function of the control parameter \( b = \sqrt{\mu K L^2 / D} \). 
  The right panel shows the corresponding minimal mean first-passage time \( \tau_{\text{sr}}^* = \tau_{\text{sr}}(\tau^*) \). 
  The red line represents the constant values obtained for the standard SR protocol, which does not depend on \( K \).}
  \label{fig:T-opt-src} 
\end{figure}

In the SR-C protocol, we introduce an additional control parameter: the trap stiffness \( K \). As a result, the value of \( \tau^* \) 
that minimizes the search time \( \tau_{\text{sr}} \), defined by
\begin{equation}
\frac{d\tau_{\text{sr}}}{d\tau} \bigg|_{\tau = \tau^*} = 0,
\end{equation}
becomes a function of \( K \). In Fig.~(\ref{fig:T-opt-src}), we plot this optimal value \( \tau^* \) as a function of the 
dimensionless control parameter \( b = L /\sqrt{D\tau_K} = \sqrt{\mu K L^2 / D} \).  
The plot on the right-hand side indicates that we can bring $\tau_{sr}$ to an arbitrary low value by increasing $K$.  
The plot on the left-hand side indicates that the corresponding optimal $\tau^*$ shifts toward larger values, 
indicating that we need an increasingly broader distribution of initial points $x_0$ (since $\tau$ controls the 
distribution $n_{\text{off}}$) with increasing $K$.  
\graphicspath{{figures/}}

\section{Conclusion}
\label{sec:sec6}

In this work, we examined a fluctuating harmonic potential as a physical realization of stochastic resetting.  
By enforcing that all return trajectories terminate at the origin, we ensured a well-defined and reproducible 
starting point for each search cycle, thereby preserving a key structural assumption of the standard SR framework.

A distinctive feature of our model is the use of \emph{information} to determine the precise moment at which 
the trap is switched off.  
This feedback shortens the duration of return trajectories without incurring additional mechanical energetic 
cost.  
In this respect, our setup falls within the class of Maxwell-demon--type models, where information is used to 
obtain an operational advantage---here, a reduction of return time.

Building on this controlled setting, we analyzed several search protocols that incorporate the return dynamics 
explicitly.  
To compute the mean first passage time (MFPT), we decomposed each cycle into statistically defined generations 
and expressed the relevant time scales and probabilities within a splitting-probability framework.

A particularly notable case is the protocol in which outward excursions are eliminated entirely and the search 
is performed solely through return motion.  
This inversion of the conventional outward-diffusion viewpoint yields an MFPT that can be written in closed form 
using the previously defined time-scale parameters.  
Under strong confinement, this return-only protocol becomes more efficient than standard SR---not because 
individual return trajectories are inherently more effective, but because they are significantly faster.

%
%

\begin{acknowledgments}
D.F. acknowledges financial support from FONDECYT through grant number 1241694.  
\end{acknowledgments}

\section{DATA AVAILABILITY}
The data that support the findings of this study are available from the corresponding author upon reasonable request.

\bibliography{harmonic-switch-v3}


\end{document}